\documentclass[fleqn,usenatbib]{mnras}

\usepackage{newtxtext,newtxmath}

\usepackage[T1]{fontenc}

\DeclareRobustCommand{\VAN}[3]{#2}
\let\VANthebibliography\thebibliography
\def\thebibliography{\DeclareRobustCommand{\VAN}[3]{##3}\VANthebibliography}

\usepackage{graphicx}	
\usepackage{amsmath}	

\usepackage{caption}
\usepackage{subcaption}
\usepackage{gensymb}
\usepackage{placeins}
\usepackage{lineno}

\title[The shocks of Cir X-1]{Dynamic shocks powered by a wide, relativistic, super-Eddington outflow launched by an accreting neutron star in the mid-20th century}

\author[F. J. Cowie et al.]{F. J. Cowie,$^{1}$\thanks{E-mail: fraser.cowie@physics.ox.ac.uk}
R. P. Fender,$^{1,2}$ I. Heywood,$^{1,3}$
F. Carotenuto,$^{4}$ J. H. Matthews,$^{1}$ B. Reville,$^{5}$ L. Olivera-Nieto,$^{6,7}$ \newauthor A. J. Cooper,$^{1}$  A. K. Hughes,$^{1}$    K. Savard,$^{1}$ P. A. Woudt,$^{2}$ J. van den Eijnden,$^{7}$ N. Grollimund,$^{8}$ P. Saikia$^{9}$
\\
$^{1}$Department of Physics, University of Oxford, Denys Wilkinson Building, Keble Road, Oxford OX1 3RH, UK \\
$^{2}$Department of Astronomy, University of Cape Town, Private Bag X3, 7701 Rondebosch, South Africa\\
$^{3}$Centre for Radio Astronomy Techniques and Technologies, Department of Physics and Electronics, Rhodes University, PO Box 94,
Makhanda, 6140, South Africa\\
$^{4}$INAF, Osservatorio Astronomico di Roma, Via Frascati 33, I-00078 Monte Porzio Catone, Italy\\
$^{5}$Max-Planck Institute for Nuclear Physics, 69117 Heidelberg, Germany\\
$^{6}$Gravitation and Astroparticle Physics Amsterdam Institute, University of Amsterdam, Science Park 904, 1098 XH Amsterdam, The Netherlands\\
$^{7}$Anton Pannekoek Institute for Astronomy, University of Amsterdam, Science Park 904, 1098 XH Amsterdam, The Netherlands\\
$^{8}$Université Paris Cité, Université Paris-Saclay, CEA, CNRS, AIM, F-91191 Gif-sur-Yvette, France\\
$^{9}$Department of Astronomy, Yale University, PO Box 208101, New Haven, CT 06520-8101, USA\\
}

\date{Accepted XXX. Received YYY; in original form ZZZ}

\pubyear{2026}

\begin{document}
\label{firstpage}
\pagerange{\pageref{firstpage}--\pageref{lastpage}}
\maketitle

\begin{abstract}

Accreting systems can launch powerful outflows which interact with the surrounding medium. We combine new radio observations of the accreting neutron star X-ray binary (XRB) Circinus X-1 (Cir X-1) with archival radio observations going back 24 years. The $\sim3$ pc scale wide-angle radio and X-ray emitting caps found around Cir X-1 are identified as synchrotron emitting shocks with significant proper motion and morphological evolution on decade timescales. Proper motion measurements of the shocks reveal they are mildly relativistic and decelerating, with apparent velocity of $0.14c\pm0.03c$ at a propagation distance of 2 pc. We demonstrate that these shocks are likely powered by a hidden relativistic ($\gtrsim0.3c$) wide-angle conical outflow launched in $1972\pm3$, in stark contrast to known structures around other XRBs formed by collimated jets over 1000s of years. The minimum time-averaged power of the outflow required to produce the observed synchrotron emission is $\sim0.1L_\text{Edd}$, while the time-averaged power required for the kinetic energy of the shocks is $\sim40 \left(\frac{n}{10^{-2} \text{cm}^{-3}}\right)L_\text{Edd}$, where $n$ is the average ambient medium number density. This reveals the outflow powering the shocks is likely significantly super-Eddington. We measure significant linear polarisation up to $52\pm6\%$ in the shocks demonstrating the presence of an ordered magnetic field of strength $\sim200~\mu\text{G}$. We show that the shocks are potential PeVatrons, capable of accelerating electrons to $\sim0.7~\text{PeV}$ and protons to $\sim20~\text{PeV}$, and we estimate the injection and energetic efficiencies of electron acceleration in the shocks. Finally, we predict that next generation gamma-ray facilities may be able to detect hadronic signatures from the shocks. 

\end{abstract}

\begin{keywords}
X-rays: binaries -- stars: neutron -- ISM: jets and outflows -- X-rays: individual: Cir X-1 -- shock waves -- acceleration of particles

\end{keywords}




\section{Introduction}



X-ray binaries (XRBs) are composed of a compact object, either a black hole (BH) or a neutron star (NS), accreting from a companion star. They offer a unique laboratory to study the physics of accretion and jets due to their evolution on human accessible timescales. X-ray binaries can deposit large fractions of their accretion energy into the ambient medium, both through wind (e.g. \citealt{ponti_2012}, \citealt{xrism_2025}) and jet outflows (e.g. \citealt{bright_2020} and \citealt{motta_2025}), and through X-ray radiation (e.g. \citealt{jeon_2014}). Much like the feedback from super-massive black holes, which is thought to play a crucial role in regulating galactic-scale processes, e.g. star formation \citep{insibashi_2012} and chemical enrichment \citep{villar_2024}, feedback from XRBs impacts the surrounding ISM in a variety of important ways. The outflows from XRBs reintroduce a fraction of the infalling gas (e.g. \citealt{zuo_2025}), energise the surrounding medium (e.g. \citealt{fender_munoz_2016}), impact its density and cause turbulence \citep{savard_2025}, and drive the production of cosmic rays (e.g. \citealt{wang_2025}). These phenomena in turn can then induce local star formation \citep{mirabel_2015}, and may introduce seed magnetic fields into the ambient medium which contribute to the magnetic fields observed in our galaxy \citep{heinz_2008}. Studying the sites where outflows interact with the ambient medium, where feedback occurs, sheds light on the process of feedback, as well as poorly understood fundamental properties of the outflows themselves, particularly their composition and energetics.

Sites where outflows interact with the ambient medium on a large ($>1~\text{pc}$) scale have only been observed in a handful of XRBs, unlike in AGN. This is likely due to the different environments which these objects inhabit \citep{heinz_2002}, and the larger sample of AGN available. In the BH XRB Cygnus X-1 (Cyg X-1), a relativistic collimated persistent jet has inflated a large ($\sim5~\text{pc}$) bubble in the ambient medium and is powering a bow shock structure (\citealt{gallo_2005}, \citealt{russell_2007}, \citealt{sell_2015}, \citealt{atri_2025}). This has allowed the energy deposited into the ambient medium by this jet to be measured, and hence provided constraints on the time-averaged power of the jet. In the BH XRB GRS~1915+105 a similar structure has recently been discovered (\citealt{tetarenko_2018}, \citealt{motta_2025}), alongside hotspot regions where strong shocks are argued to accelerate particles \citep{kaiser_2004}. In both of Cyg X-1 and GRS 1915, it is thought to be the collimated hard state jet powering these structures, which is active for the majority of the time (e.g. \citealt{grinberg_2013}). This has allowed for models originally developed for AGN lobes to be applied to understand these interaction sites and use them as calorimeters for the overall jet power (\citealt{kaiser_1997}, \citealt{kaiser_2004}). Hard state jets are also thought to power the large scale lobes ($\sim4~\text{pc}$ seen in GRS 1758-258 (\citealt{sunyaev_1991}, \citealt{marti_2002}, \citealt{tetarenko_2020}, \citealt{mariani_2025}), which appear to show hydrodynamic backflow features \citep{marti_2017}, and the large scale ($\sim2~\text{pc}$) lobes in 1E 1740.7-2942 (\citealt{sunyaev_1991}, \citealt{mirabel_1992}, \citealt{luque_2015}). A complex system showing an array of outflow interactions with the ambient medium is SS 433 \citep{margon_1984}. Outflows have inflated bubbles in the nebula surrounding SS 433, thought to be its natal supernova remnant \citep{downes_1986}. Furthermore, radio filaments \citep{goodall_2011}, and X-ray knots on scales $>30~\text{pc}$ \citep{safi_harb_2022} may be be regions where the precessing jet or another outflow in the system interacts with the ambient medium. 

Very large scale structures ($>100~\text{pc}$) are also found in some extra-galactic XRBs, such as S26 in NGC 7793 (\citealt{pakull_2010}, \citealt{soria_2010}), LMC X-1 \citep{hyde_2017}, and S10 in NGC 300 \citep{urquhart_2019}. Large scale radio, X-ray and gamma ray structures have also been found around V4641 Sgr (\citealt{alfaro_2024}, \citealt{suzuki_2025}, \citealt{hess_v4641}, Grollimund et al., in press). Finally, resolved gamma-ray structures have recently been found around a subset of XRBs \citep{lhaaso_2024}.

\subsection{Circinus X-1}

Circinus X-1 (Cir X-1) \citep{margon_1971} is the only confirmed neutron star (\citealt{tennant_1986}, \citealt{linares_2010}) X-ray binary (NSXB) to show surrounding interaction-driven large scale structure. Early observations revealed a $\sim10~\text{pc}$ circular radio nebula (\citealt{haynes_1986}, \citealt{tudose_2006}), initially thought to be a cavity inflated by the jets of this source. However, further X-ray observations revealed the X-ray counterpart to this nebula as the natal supernova remnant (SNR) of Cir X-1 \citep{heinz_2013}, and higher resolution radio observations revealed an edge brightened morphology consistent with a young SNR \citep{calvelo_2012a}. These findings make Cir X-1 the youngest known XRB, with an age of $2900$~years ($<5400$ years at 3-sigma) \citep{heinz_2013}. Despite the fact that the $\sim10~\text{pc}$ radio nebula, known as the \emph{Africa nebula} (see Figure~\ref{fig:whole_nebula}), was not inflated by accretion powered outflows from the source, improved observations showed the presence of $\sim3~\text{pc}$ wide-angled radio and X-ray cap structures within the SNR, in the NW and SE directions. These caps were inferred to be shocks where an accretion powered outflow interacts with the ambient medium \citep{sell_2010}. Furthermore, recent radio observations with MeerKAT revealed the presence of asymmetric bubbles inflated in the SNR, and modelling indicates that these were formed by a powerful collimated jet early in the evolution of the system (\textcolor{blue}{Gasealahwe, Savard et al. 2025})\nocite{gasealahwe_2025}. Furthermore, higher resolution MeerKAT observations have revealed a collimated precessing jet in the system which propagates out to at least $\sim1~\text{pc}$ scales \citep{cowie_2025}. 

Cir X-1 has been monitored and studied extensively across the electromagnetic spectrum over the 50+ years since its discovery \citep{margon_1971}. It famously shows a $\sim16.6$ day period lightcurve at radio, infrared, optical and X-ray wavelengths (\citealt{calvelo_2012a}, \citealt{glass_1978}, \citealt{johnston_2016}, \citealt{tominaga_2023}). This behaviour is best explained by an eccentric binary model where the neutron star accretes material from a companion star, where the emission is modulated by a combination of variable mass transfer rate and absorption along the line of sight (\citealt{schulz_2020}, \citealt{tominaga_2023}, \citealt{yu_2024}, \citealt{masahiro_2025}) throughout the eccentric orbit (\citealt{murdin_1980}, \citealt{johnston_1999}). However, the details of this model, along with the nature of the companion star and the orbital parameters of the system, remain uncertain \citep{moneti_1992}. Observations point to either a high mass companion (\citealt{jonker_2007}, \citealt{schulz_2020}) or a non-standard low mass companion whose evolution is affected by the supernova or binary evolution \citep{johnston_2016}. The detection of Type \MakeUppercase{\romannumeral 1} X-ray bursts confirmed the neutron star nature of the compact object (\citealt{tennant_1986}, \citealt{linares_2010}, \citealt{yu_2025}). The X-ray spectral and timing properties of Cir X-1 are sometimes similar to low mass NSXBs, exhibiting both Z and Atoll behaviour \citep{soleri_2009}, and QPOs \citep{boutloukos_2006}. Cir X-1 has also been observed to have an evolving X-ray polarisation angle over the 16.6 day orbit \citep{rankin_2024}. The distance to the system is well constrained to be $\sim9.4~\text{kpc}$ (see \citealt{heinz_2013} and references therein for a discussion on the distance to the source). This distance measurement implies that Cir X-1 sometimes exceeds the Eddington limit for a $1.4 \: \textup{M}_\odot$ neutron star. 

A transient relativistic jet is launched by Cir X-1 near periastron \citep{moin_2011}, when the accretion rate is presumably highest, and this leads to a radio flare consistent with an expanding synchrotron-emitting plasma (\citealt{vdL}, \citealt{haynes_1978}, \citealt{calvelo_2012a}). The radio flux density of these flares has varied over several orders of magnitude from sub-mJy to multiple Jy (at $\sim1~\text{GHz}$) over decade timescales \citep{armstrong_2013}. The jets of Cir X-1 have also been resolved and studied for over 20 years, on various scales. VLBI observations showed a well collimated jet in a NW-SE direction \citep{miller-jones_2011b}. On larger scales the jets have been observed to have different position angles with distance from the core \citep{coriat_2019}. \cite{cowie_2025} showed the jets had a position angle in May 2025 in a NE-SW direction, and that the jets have a curved radio morphology on $\sim1~\text{pc}$ scales. Furthermore, Over 20 years the jet position angle has varied over a range of $>110\degree$. All of these observations are consistent with precessing jets launched by Cir X-1. The presence of an unseen ultra-relativistic flow has been inferred from core-lobe time-delays in the SE direction \citep{fender_2004}, but the recurrent flaring of Cir X-1 potentially complicates this interpretation. Finally, the presence of a powerful and variable accretion disc wind has been inferred from P Cygni profiles in X-ray spectra (\citealt{brandt_2000}, \citealt{schulz_2002})

Overall, while Cir X-1 remains a definitively peculiar XRB, it is a unique laboratory to study the early stages of evolution of XRBs. Its confirmation as a neutron star accretor and launching of powerful observable outflows allow comparisons to BH XRBs in order to understand differences between these objects. In our case, the fact that it is the only confirmed neutron star XRB to show large scale structures where outflows interact with the ambient medium makes it a truly invaluable object to study feedback from an accreting neutron star. 

In this work we present observations spanning over two decades of the large scale ($>1~\text{pc}$) structures surrounding Cir X-1 within the \emph{Africa nebula}, with the aim of studying their time evolution. We focus on the evolution, nature, properties, and energetics of the radio and X-ray cap features, discovered by \cite{sell_2010}. Section~\ref{sec:obs} describes the observations and the data reduction process. Section~\ref{sec:results} shows the results of the observations, detailing the evolution of the radio/X-ray caps over two decades. Section~\ref{sec:discussion} discusses the implication of the observed evolution of the caps. In particular the nature of the caps (\ref{sec:cap_nature}), geometry of the powering outflow (\ref{sec:geometry}), energetics (\ref{sec:energetics}), and particle acceleration properties (\ref{sec:particle_accel}), are analysed. Finally, Section~\ref{sec:comparison} compares the caps to other known large scale structures around XRBs and Section~\ref{sec:outflow_nature} discusses the possible nature of the remarkable powering outflow.



\section{Observations \& data processing}\label{sec:obs}

In this work we have used radio observations of Cir X-1 and its caps, spanning 24 years, at several frequencies, with two different connected element interferometers. This results in 5 epochs of observation between 2001 and 2025. Below we detail our data reduction, analysis and imaging procedure.

\subsection{ATCA Observations}

For the data at earlier times (2011 and before) we reanalysed archival observations from the Australia Telescope Compact Array (ATCA). All of these data were obtained from the Australia Telescope Online Archive. We used $\sim9$ hours of consecutive on source time with ATCA at 1.4 and 2.5~GHz, with observations starting on 03/08/2001. J1934-638 was observed for 17 minutes as a primary calibrator and $\sim15$ minute observations of Cir X-1 were sandwiched by 2 minute observations of J1520-58 to use as a secondary calibrator. These observations were taken with the original 2x128~MHz ATCA correlator and with the 1.5A array configuration. These data were originally presented in \cite{tudose_2006} and taken under project code C917 (PI: Fender). We also used $\sim19$ hours of consecutive on source time with ATCA at 2.1~GHz, with observations starting on 15/12/2011. J1934-638 was observed for 20 minutes as a primary calibrator and 20 minute observations of Cir X-1 were sandwiched by 2 minute observations of J1511-55 to use as a secondary calibrator. These observations were taken with a bandwidth of 2048~MHz using the Compact Array Broadband Backend (CABB) \citep{cabb}, and in the 6A array configuration. These data were originally presented in \cite{coriat_2019} and taken under project code C2597 (PI: Calvelo).

Both the 2001 and 2011 ATCA data were flagged for radio frequency interference (RFI) and calibrated using standard techniques within \textsc{casa} \citep{casa}. The 2001 ATCA data were then imaged using \textsc{wsclean} \citep{wsclean} with a Briggs weighting of 0 \citep{briggs}. 

Initial imaging of the 2011 ATCA data was done using 6 frequency channel multi-frequency deconvolution in \textsc{wsclean}. From these initial images a deconvolution mask was made using \textsc{breizorro} \citep{breizorro} and used in a subsequent round of imaging with a deeper cleaning threshold. The resulting improved sky model was then used for a round of self calibration using \textsc{casa}. Significant artefacts remained after this process around Cir X-1 and bright sources far from the primary beam centre, as noted by \cite{coriat_2019}. In the case of Cir X-1 these artefacts are likely because the source is variable in flux density over the observation. We adopt the same approach as \cite{coriat_2019} to deal with these artefacts, using the \textsc{DDFacet} \citep{ddf} imager combined with the \textsc{killms} \citep{killms} gain solver to obtain and apply calibration solutions for 7 different tessels on the sky, the same tessels as used in \cite{coriat_2019}. A Briggs weighting of 0.5 was used throughout this process.

\subsection{MeerKAT Observations}

For the later data (2018 and later) we used observations of Cir X-1 and the surrounding field using the MeerKAT radio telescope \citep{meerkat}. These observations were taken as part of the ThunderKAT \citep{thunderkat}, and X-KAT programmes (MeerKAT Proposal ID: SCI-20230907-RF-01). We used a 45 minute L-band (1.3~GHz central frequency, 856~MHz bandwidwth) observation taken on 27/10/2018, a 240 minute S2-band (2.6~GHz, 875~MHz bandwidth, \citealt{s_band}) observation taken on 08/07/2023, and a 60 minute S2-band observation taken on the 31/05/2025. The length of the observation refers to the total on source time, composed of scans up to 30 minutes long sandwiched by 2 min secondary calibrator (J1427–4206) scans. Each observation began and ended with a 5-minute scan of J1939–6342, the primary and polarisation leakage calibrator, and included a 10 minute observation of J1331+3030 to be used as a polarisation angle calibrator.

All of the MeerKAT observational data were reduced using the semi-automated  pipeline \textsc{polkat}\footnote{\url{https://github.com/AKHughes1994/polkat}} \citep{polkat} (based on the \textsc{oxkat} software, \citealt{oxkat}). The software packages used in this pipeline were accessed using \textsc{singularity} for software containerisation \citep{singularity}. First generation calibration, including polarisation calibration (applying corrections from the primary, secondary and polarisation calibrators, see \citealt{heywood_2022}) and iterative RFI flagging was done using \textsc{casa} \citep{casa}. The target field was flagged further using \textsc{tricolour} \citep{tricolour} and then imaged using \textsc{wsclean} \citep{wsclean}. A deconvolution mask was manually made and iteratively improved through further imaging and use of the mask making tool \textsc{breizorro} \citep{breizorro}, before second generation calibration (phase and delay self calibration) was performed using \textsc{quartical} \citep{quartical}. Joined frequency channel and joined polarisation deconvolution using 8 channels and a Briggs weighting of 0.0 was performed. The astrometry of the images as an ensemble was checked using a selection of 10 background sources in the field of varying flux density. No systematic offset or significant scatter ($>10\%$ of the synthesised beam) was found.


\section{Results}\label{sec:results}






\begin{figure*}
 \includegraphics[width=1.5\columnwidth]{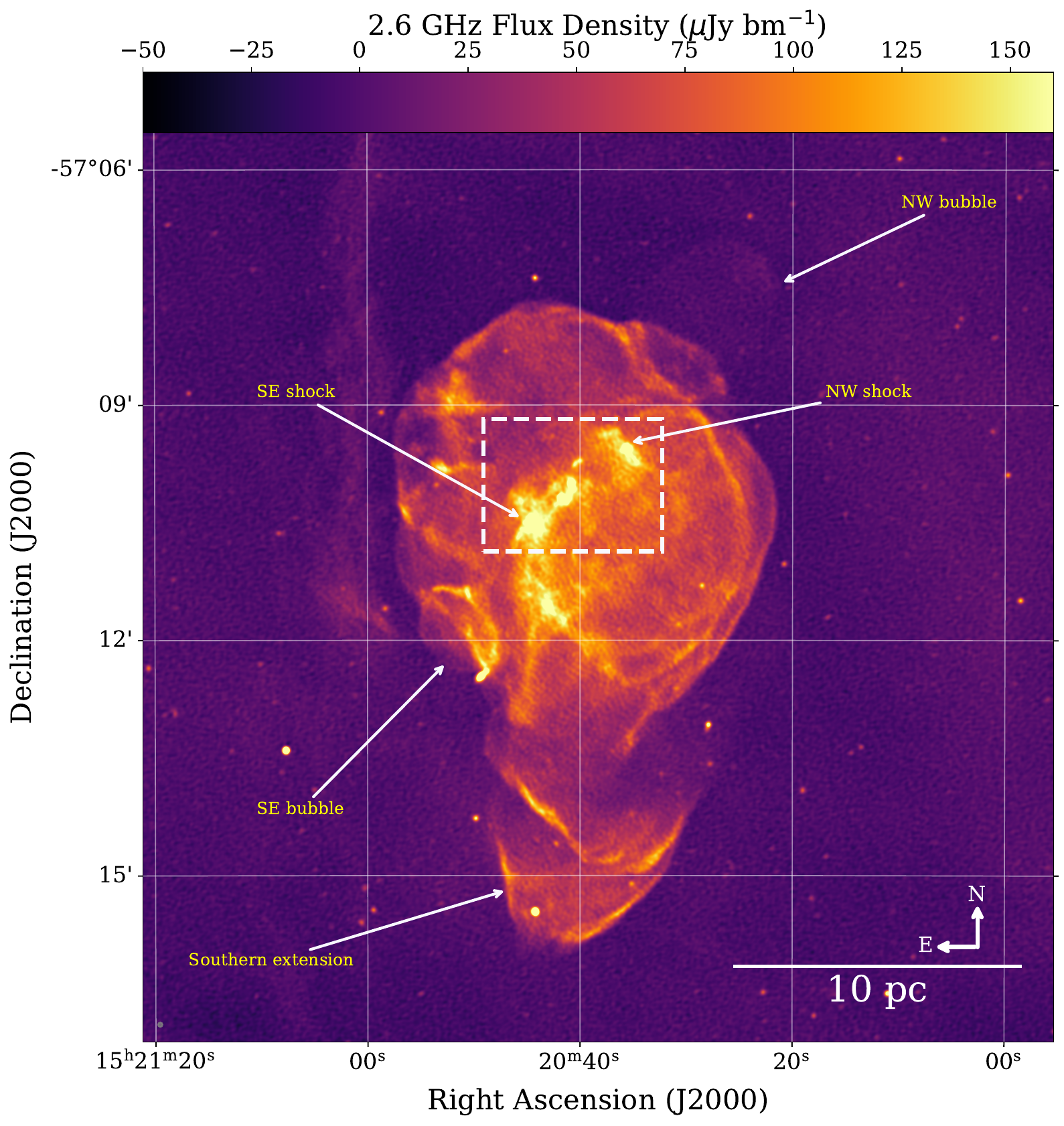}
 \caption{2.6~GHz MeerKAT image in 2023 of Cir X-1 and its natal SNR, the \emph{Africa nebula}. Structures of radio emission associated with the SNR and areas where the outflows from Cir X-1 interact with the surrounding region are visible and labelled. Cir X-1 and its precessing jets on parsec scales are seen as radio sources in the centre of the nebula. To the north-west and south-east asymmetric jet inflated bubbles are visible at the edge of the SNR. Roughly half way between Cir X-1 and the edge of the SNR the radio cap structures are seen to the north-west and south-east. An unlabelled version of this plot is available in the online supplementary material.}
\label{fig:whole_nebula}
\end{figure*}

\begin{figure*}
 \includegraphics[width=2.0\columnwidth]{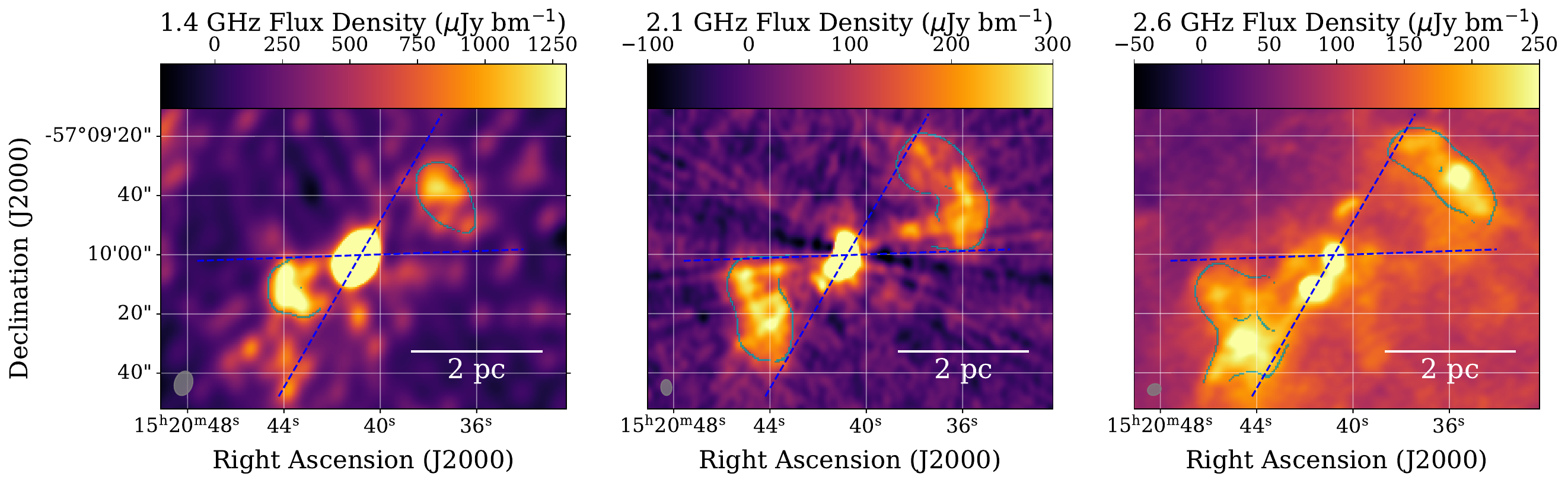}
 \caption{Three images showing Cir X-1 and the radio caps in 2001 at 1.4 GHz, 2011 at 2.1 GHz, and 2023 at 2.6 GHz. Evolution in the morphology and position of the radio caps over the different epochs is evident. Despite this, the straight dashed lines represent an opening angle of $29\degree$, illustrating the constant opening angle over the three epochs. Edges found by the gradient based edge detection algorithm described in text are shown, and the beam size is illustrated by an ellipse in the lower left. Despite the fact that different interferometer sampling can cause apparent differences in images of diffuse structure we are confident in the structural and positional changes observed (see text for more). A .gif showing the evolution of the caps over all 5 epochs is provided in the online supplementary material.}
 \label{fig:triple_figure}
\end{figure*}

\begin{figure}
 \includegraphics[width=\columnwidth]{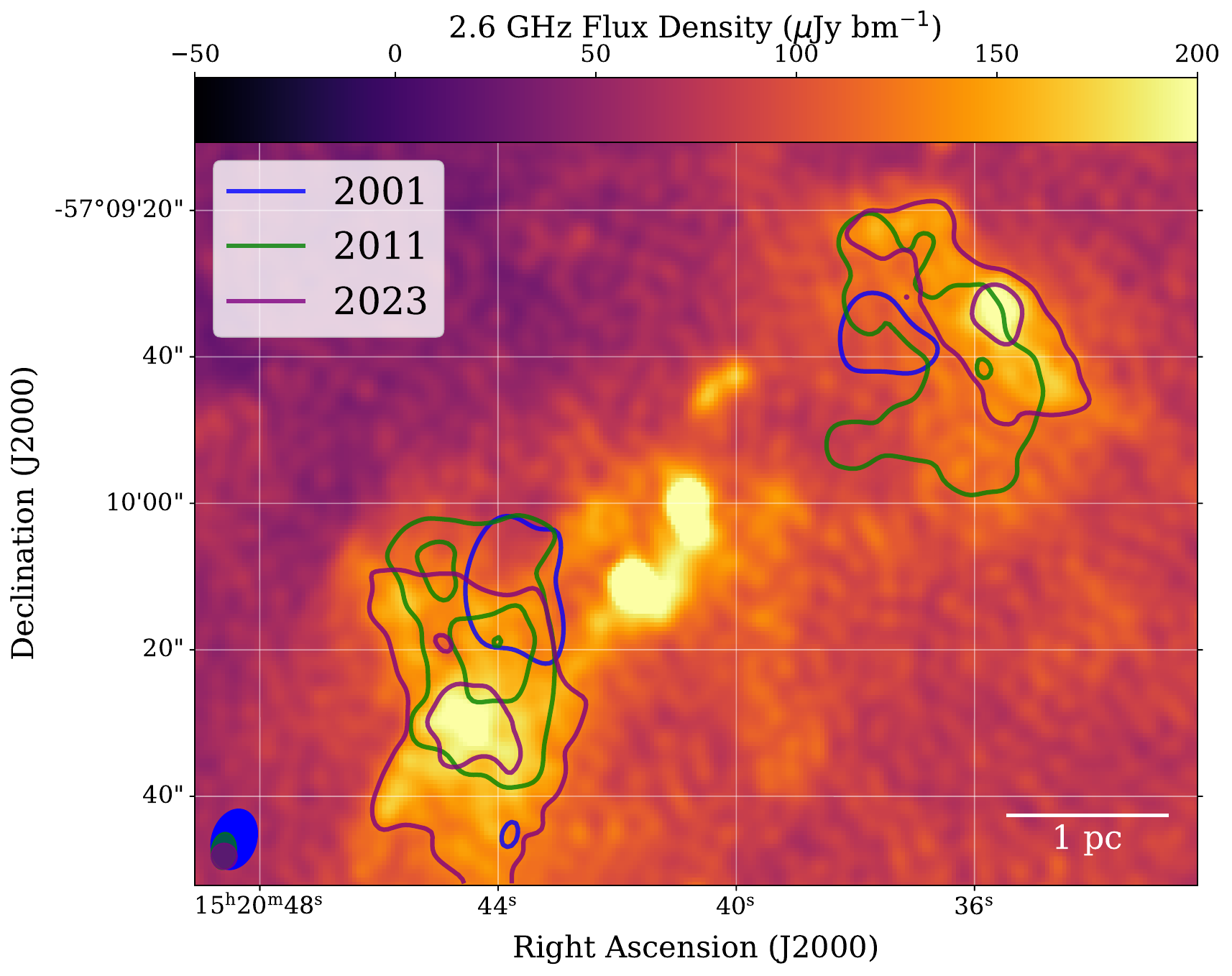}
 \caption{Image of Cir X-1 and the radio caps in 2023 with 6 and 12 $\sigma$ contours from the 2001, 2011, and 2023 images over-plotted, clearly showing the motion of the radio caps away from Cir X-1. The position angle of the brightest region in the caps, the hotspot, also varies over the 3 epochs.  The beams of all 3 observations are shown in the bottom left.}
 \label{fig:shock_contours}
\end{figure}

\begin{figure*}
 \includegraphics[width=1.6\columnwidth]{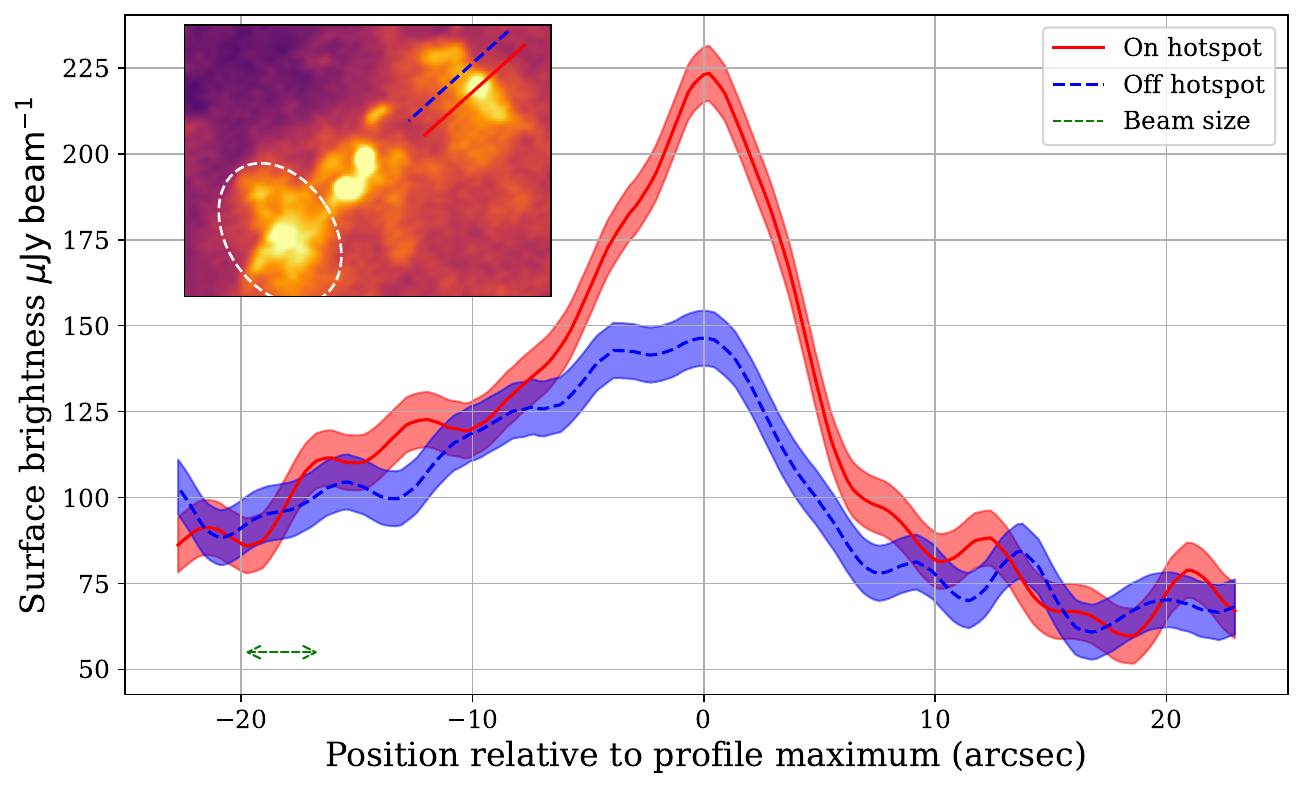}
 \caption{Surface brightness profiles of the NW cap along a position angle of $-48\degree$ E of N, perpendicular to the long axis of the caps. In one case the surface brightness profile crosses the hotspot and in the other case it does not. The lines along which the profiles are measured are shown in an inset in the top left in addition to an example of the ellipse used for flux density measurements discussed in text. The surface brightness profiles are aligned to their maximum for ease of comparison. An asymmetry in both profiles is apparent, with a sharp drop in surface brightness on the side of the cap furthest from Cir X-1, expected in the case of a shock. The shaded region represents the uncertainties on surface brightness measurements from the image noise. The size of the beam, the expected size scale of correlated image noise, is shown in the bottom left by the double headed arrow.}
 \label{fig:flux_cut}
\end{figure*}

\begin{figure*}
 \includegraphics[width=1.9\columnwidth]{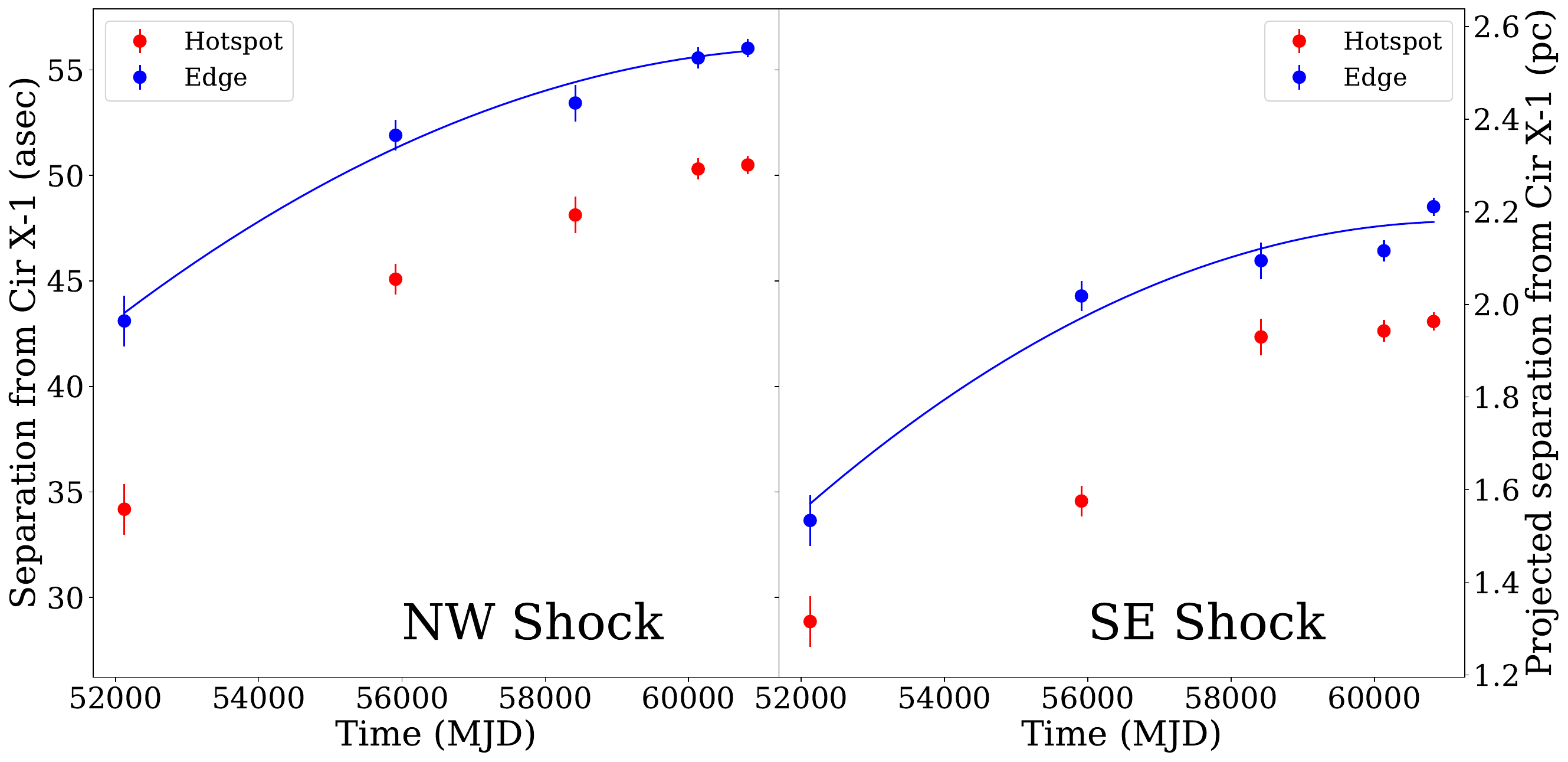}
 \caption{Separation of the caps from Cir X-1 as a function of time for the NW edge and hotspot (left), and the SE edge and hotspot (right). For the edge positions the best fit deceleration model is shown, with parameters given in Table~\ref{tab:position_fits}. An asymmetry in the separation from Cir X-1 between the NW and SE caps and evidence for deceleration in both cases is evident.}
 \label{fig:position_curves}
\end{figure*}

\subsection{Imaging and flux density measurements}

Figure~\ref{fig:whole_nebula} shows the result of the 2023 S2-band observation of Cir X-1 and its natal SNR, known as the \emph{Africa nebula} (\textcolor{blue}{Gasealahwe, Savard et al. 2025}). This is the most detailed image of Cir X-1 and the \emph{Africa nebula} to date with a resolution of $3.5''$ and a noise level of $4~\mu\text{Jy}$ using a Briggs weighting of $-0.3$. Showcased in the image are several striking features of the SNR including the inflated bubbles and associated rings from previous jet activity (\textcolor{blue}{Gasealahwe, Savard et al. 2025}), the radio/X-ray caps along the historic jet axis in the source (see \citealt{cowie_2025} for a discussion on the variable jet axis), the double shock structure of the SNR shell \citep{calvelo_2012a}, and the southern extension to the nebula of unknown origin.

Figure~\ref{fig:triple_figure} shows the region of interest in the SNR for this work, containing the radio/X-ray caps, at 3 different epochs. From left to right, an ATCA image at 1.4~GHz from 2001 with a resolution of $8.4''$ and a noise level of $120~\mu\text{Jy}$ using a Briggs weighting of 0; an ATCA image at 2.1~GHz from 2011 with a resolution of $5.3''$ and a noise level of $12~\mu\text{Jy}$ using a Briggs weighting of 0.5; and the zoom in of the 2.6~GHz MeerKAT image shown in Figure~\ref{fig:whole_nebula} from 2023. The images all show a central radio source, Cir X-1, of varying brightness, and are not flux matched due to the different observing frequencies, noise levels, and uv-coverage. In each observation the radio/X-ray caps are clearly detected above the noise and the most important result of this work can be seen. There is clear evolution in the morphology and position of the caps over a timescale of $\sim20$ years. 

To better demonstrate this, Figure~\ref{fig:shock_contours} shows the 2023 MeerKAT image, overlaid by contours at 6 and 12 $\sigma$ levels for the 2001, 2011, and 2023 images. Contours from the region immediately surrounding the central source, Cir X-1, have been removed for clarity. Figure~\ref{fig:shock_contours} shows that both the north-west (NW) cap and the south-east (SE) cap show proper motion away from Cir X-1 over this time period. We note that the proper motion limit for Cir X-1 of $<5~\text{mas yr}^{-1}$ means that Cir X-1 itself will have no significant proper motion in our observations over 24 years \citep{mignani_2002}. In the higher resolution images in particular, we observe that the caps do not have a completely smooth morphology, instead there appears to be a single bright spot, or hotspot. We observe that the position angle of this hotspot, measured from the Cir X-1, changes over time, by at least $15\degree$. This evolution is most clear from 2001 to 2011 in the NW cap, and between 2011 and 2023 in the SE cap. We also observe that in general, the hotspot structures in the NW and SE caps in a given epoch are not symmetrical about the central source, Cir X-1. We measure the half opening angle of each cap at each epoch from Cir X-1 and find a tight range of values from $26\degree$ to $29\degree$, within measurement uncertainties, demonstrating the opening angle remains approximately constant over time despite other evolution, and is the same for the NW and SE caps. 

While different interferometer sampling can cause apparent differences in images of diffuse structure, we do not observe any evidence that this is responsible for the proper motion or evolution of the shock structures in our observations. We find proper motion even comparing observations with similar uv-coverage (e.g. MeerKAT 2018 and MeerKAT 2025), and we find no significant proper motion or evolution in other diffuse structures (e.g. the shock at the edge of the SNR) throughout the \emph{Africa nebula} over our observations. Furthermore, while we are comparing observations at different frequencies we see no evidence for any significant frequency dependent morphology (over our observed range of frequencies). For example observations taken closer together in time at different frequencies show similar morphology (e.g. MeerKAT 2018 and MeerKAT 2023). Therefore, we are confident in the structural and positional changes in our observations.





We examine the surface brightness profile perpendicular to the long axis of the NW cap, at a position angle of $-48\degree$ E of N. Figure~\ref{fig:flux_cut} shows the surface brightness profile across the NW cap in the 2023 image, intersecting the hotspot, and over a different region. The lines over which the profiles are measured are shown in the inset of Figure~\ref{fig:flux_cut}. The profiles are aligned relative to their spatial maxima for easy comparison, and a negative displacement corresponds to the direction towards Cir X-1. From Figure~\ref{fig:flux_cut} it is apparent that the surface brightness profile of the caps perpendicular to their long axis is asymmetric. They can be described as having a slow rise to a maximum on the side closest to Cir X-1, followed by a steep drop to the background emission level on the side of the cap furthest from Cir X-1. This morphology is present in both caps. From the profiles we measure the extent of the caps in the direction towards Cir X-1 to be $\sim30''$, or a projected distance of $\sim1.4 \: \text{pc}$, where we have used the distance to the source of $9.4~\text{kpc}$ \citep{heinz_2015}, which is adopted hereafter.

Measuring the total flux density of the caps is not straightforward given the presence of non-uniform background radio emission from the SNR. Using the MeerKAT 2023 2.6 GHz image we measure the flux density, integrated over an ellipsoidal region encompassing the cap emission, to be $15\pm0.2 \:\text{mJy}$ and $17\pm0.2 \: \text{mJy}$ for the NW and SE cap respectively. This ellipsoidal region, which is the same size for the NW and SE caps, is shown for the SE cap in the inset of Figure~\ref{fig:flux_cut}. The measured flux density however will include a significant contribution from the radio emission from the SNR. To estimate the SNR contribution we measure the flux density integrated over the same ellipsoidal region at two positions in the SNR with no prominent features and which are at the extreme ends of background emission. These regions have integrated flux densities of $7 \: \text{mJy}$ and $11 \: \text{mJy}$. By subtracting this from the measured flux density of the caps, we obtain values for the flux density of the caps of $6\pm2 \: \text{mJy}$ and $8\pm2 \: \text{mJy}$ for the NW and SE cap respectively, where the conservative uncertainty is dominated by the unknown background contribution. The flux density of the caps is equal within these uncertainties.

Comparison to flux densities measured at previous epochs is challenging to do accurately given the different observing frequencies and instruments. We therefore do not attempt to make any statement on the total flux density evolution between epochs of the shocks. However, within each observation we find no evidence for 
significant flux asymmetry between the two caps in any epoch. We note that MeerKAT has a substantially improved uv-coverage when compared to ATCA and is a core-dominated array. Both of these factors make it more sensitive and capable for measurements of extended emission and so we use the MeerKAT measurements as reported above for further analysis in Section~\ref{sec:discussion}.

\subsection{Polarisation properties}

In the 2023 MeerKAT image, we can also investigate the polarisation properties of the caps. We use 32-channel multi-frequency, joined polarisation deconvolution in \textsc{wsclean} \citep{wsclean} to create images for all Stokes parameters, and for this specific imaging run to investigate polarisation we use an inner Tukey taper from 0 to $10^4 \lambda$ (where $\lambda$ is the observing wavelength) to down-weight short baselines and limit background emission from the nebula. This improves the noise level inside the SNR. We then use the resolution homogenisation function of \textsc{polkat} to ensure our channelised images for each Stokes parameter are homogeneous in resolution across the frequency range. \textsc{rm-tools} \citep{rmtools} is then used to run a 3D rotation measure synthesis \citep{brentjens_2005} analysis on the image. This produces a 3D polarisation intensity cube, with 2 spatial dimensions, and a 3rd dimension of Faraday depth. This analysis reveals several areas with significant linear polarisation in the \emph{Africa nebula}. A full polarisation intensity map is shown in Appendix~\ref{sec:pol_inten}. In particular, we find linear polarisation emission from both the NW and SE caps. Linear polarisation only appears in parts of the cap structures, preferentially towards the edge of the caps, and is absent from the hotspot features, down to a $5\sigma$ limit on fractional polarisation of $\sim5\%$. We measure a linearly polarised flux density of up to $30\pm3 \:\mu \text{Jy}$, a $10\sigma$ detection. We measure linear polarisation fractions of up to $40\pm4\%$ and $52\pm6\%$ in the NW and SE caps respectively. Rotation measure measurements of the caps using the polarised emission regions are consistent with $0 \:\text{rad m}^{-2}$, with an uncertainty of $\sim100 \:\text{rad m}^{-2}$. No significant circular polarisation is detected in any features in the image.

\subsection{Proper motion measurements}

\begin{table*}
 \caption{Fit parameters for the ballistic and constant deceleration models for both the NW and SE cap edges and hotspots.}
 \label{tab:position_fits}
 \begin{tabular*}{1.56\columnwidth}{@{}l@{\hspace*{24pt}}l@{\hspace*{24pt}}l@{\hspace*{24pt}}l@{\hspace*{24pt}}l@{}}
  \hline
  Parameter & NW Edge & SE Edge & NW Hotspot & SE Hotspot\\
  \hline
  $v_{\text{bal}}$ & $0.067c\pm0.006c$ & $0.070c\pm0.006c$ & $0.085c\pm0.006c$ & $0.090c\pm0.006c$\\[2pt] 
  $t_{0,\text{bal}}$ (year) & $1900\pm10$ & $1922\pm7$ & $1936\pm5$ & $1953\pm4$\\[2pt]
  $\chi^2_{r,\text{bal}}$ & 3.3 & 7.7 & 5.3 & 4.3\\[2pt]
  $v_0$ & $0.29c\pm0.06c$ & $0.30c\pm0.05c$ & $0.30c\pm0.04c$ & $0.20c\pm0.04c$\\[2pt]
  $t_{0,\text{dec}}$ (year) & $1972\pm3$ & $1979\pm3$ & $1986\pm7$ & $1976\pm4$\\[2pt]
  $a$ $\left(\times10^{-3}~\text{cm s}^{-2}\right)$ & $-4.9\pm1.8$ & $-6.0\pm1.8$ & $-6.8\pm1.8$ & $-2.5\pm1.8$\\[2pt]
  $\chi^2_{r,\text{dec}}$ & 1.1 & 5.7 & 0.4 & 5.5 \\[2pt]
  $v_{\text{2 pc}}$ & $0.14c\pm0.03c$ & $0.092c\pm0.014c$ & $0.14c\pm0.02c$ & $0.07c\pm0.02c$\\[2pt]
  $t_{\text{2 pc}}$ (year)& $2002\pm1.4$ & $2013\pm2$ & $2008\pm2$ & $2025\pm5$\\[2pt]
  \hline
 \end{tabular*}
\end{table*}

Finally, we can measure the proper motion of the caps, which can be seen by eye from Figures~\ref{fig:triple_figure} and \ref{fig:shock_contours}. Due to the complexity of the cap geometry we adopt two approaches to measure the positions of the caps. For the first of these we can find the maximum pixel within the region of the cap (which will correspond to the hotspot position) and measure the separation of this pixel from Cir X-1. Secondly, we can exploit the asymmetric nature of the flux density profile of the caps, shown in Figure~\ref{fig:flux_cut}, and find the steep "edge" of each cap and use this to define the position of the shock. This is done using a technique known as Canny edge detection \citep{canny2}, implemented using the \textsc{scikit-image} package \citep{scikit_image}. First a Gaussian filter is applied to smooth the image and reduce the impact of noise, in our case a smoothing scale of $6''$ is used. The algorithm then finds the edges using gradient based edge detection. We choose an upper and lower threshold for the algorithm of $1 \: \mu \text{Jy pixel}^{-1}$ and $10 \: \mu \text{Jy pixel}^{-1}$ respectively. We note that in general, the position of the edge found for the cap is insensitive to changes in these parameters. Finally, after inspecting the images to ascertain whether an edge has been found corresponding to the each cap, we measure the separation of the edge from Cir X-1 at a position angle of $128\degree$, chosen as this line intersects the centre of the both the NW and SE caps. Uncertainties for each of these methods are taken to be $15\%$ of the FWHM of the major axis of the restoring beam \citep{cowie_2025}. We use each of these two measurement methods for all 5 of our observing epochs (2001, 2011, 2018, 2023, 2025) as described in Section~\ref{sec:obs}.


Figure~\ref{fig:position_curves} shows the separation of the NW (left panel) and SE (right panel) caps from Cir X-1 as a function of time. The edge detection method shows a greater separation at each time from Cir X-1 than the hotspot position, as expected. There is a significant asymmetry in the separation of the NW and SE caps from Cir X-1, with the NW cap being further from Cir X-1 at all times. The separation against time curves also reveal evidence of deceleration in the proper motion of the caps. We can measure the proper motion of the caps by fitting both a ballistic (constant apparent velocity $v_\text{bal}$) model, and a model of constant apparent deceleration, $a$, to the data. The fit is performed using the \textsc{curve\_fit} module of the \textsc{scipy} package \citep{scipy} and the hotspot and edge data for each cap is treated separately, leading to a total of 4 datasets to be fit. Figure~\ref{fig:position_curves} shows the deceleration model fit to the edge positions of the caps.

The model fit degrees of freedom are 3 and 2 for the ballistic and deceleration models respectively. The likelihood-ratio chi-squared test \citep{wilks_1938} shows that in 3 of the 4 cases the deceleration model is significantly favoured ($p$-value < 0.01) over the ballistic model, with the exception being the hotspot positions for the SE cap. The fit parameters for each case are shown in Table~\ref{tab:position_fits}. For the deceleration fits we provide the apparent velocity and time extrapolated to when the caps were co-spatial with Cir X-1 ($v_0$ and $t_{0,\text{dec}}$), and also the apparent velocity and time after they have travelled a distance of 2~pc ($v_{2~\text{pc}}$ and $t_{2~\text{pc}}$). The uncertainties on the calculated time were found using a Monte Carlo analysis and the uncertainties from the fitted parameters which were: the apparent velocity at a reference time, the constant apparent deceleration, and the position at a reference time. The deceleration model is also especially preferred in modelling the edge separation given the observed symmetry in the apparent launch velocities and times for the caps. This is expected from physical considerations but not enforced as the NW and SE caps are fit separately.

In summary, from the deceleration models, we find that the caps were likely "launched" from Cir X-1 around 1975, remarkably $\sim5$ years after its discovery \citep{margon_1971}. The caps were launched with a mildly relativistic apparent velocity of $\sim0.3c$ and have since decelerated. After travelling 2~pc, there is an asymmetry in the separation and apparent velocity of the NW and SE caps, where the NW cap has travelled further and maintained a higher apparent velocity than the SE cap. This is consistent with the SE cap suffering from larger deceleration. At a distance of 2~pc the apparent velocity of the caps has reduced to $\sim0.1c$, and the caps reached this distance in the early 2000s/2010s in the NW and SE cases respectively. 

\section{Discussion}\label{sec:discussion}

Given the observational results presented in the previous section, we can explore the physical nature of the caps, their relation to Cir X-1, and measure their physical properties. Throughout, we adopt centimetre-gram-second units.

\subsection{Physical nature of the caps}\label{sec:cap_nature}




\cite{sell_2010}, who first discovered the cap structures, postulated that they were related to Cir X-1 rather than features of emission from the SNR. They argued this because of their symmetry around Cir X-1, and their different X-ray colour when compared to the emission from the SNR. We can now say these features are certainly related to the central neutron star X-ray binary rather than the SNR, given their inferred lifetime (the time elapsed since they were co-spatial with Cir X-1) of $\lesssim100$ years, compared to the $\sim2900$ year inferred age of the SNR \citep{heinz_2013}.

\cite{sell_2010} also concluded that these cap features are shock fronts powered by an outflow launched by Cir X-1 interacting with the ambient medium. This followed from the observed limb brightened morphology in X-rays and a sharp drop in surface brightness in both X-rays and radio on the sides of the caps furthest from Cir X-1. This interpretation is confirmed by the steep drop in surface brightness seen in our higher quality radio data, shown in Figure~\ref{fig:flux_cut}, and our observation of motion away from Cir X-1, as expected for shocks powered by an outflow from the binary system. Hereafter we refer to the cap structures as shocks, and we will continue to find numerous pieces of evidence which support this interpretation throughout our analysis.

The radio emission from the shocks is due to synchrotron radiation from a population of ultra-relativistic electrons. This is evident from the steep spectral index, $\alpha < -0.5$ (where $F_\nu \propto \nu^\alpha$), in the shock regions (see Figure~6 in \citealt{tudose_2006} and Figure~4 in \textcolor{blue}{Gasealahwe, Savard et al. 2025}), which is consistent with optically thin synchrotron radiation. Furthermore, the detection of a high degree of linear polarisation in the radio emission of the shock region supports a synchrotron origin. Optically thin synchrotron radiation is intrinsically polarised in a direction perpendicular to projection of the magnetic field onto the plane of the sky \citep{pacholczyk} (generally only in the absence of bulk relativistic motion e.g. \citealt{lyutikov_2005}). For a power law distribution of ultra-relativistic particles, with power law exponent $-p$ producing optically thin synchrotron radiation, the maximum degree of linear polarisation produced is $\Pi_\text{max}=\frac{p+1}{p+\frac{7}{3}}$ \citep{ginzburg_1965}. For $p\approx2$, as expected from shock acceleration of particles (see \citealt{matthews_2020} for a review), this is $\sim70\%$. As outlined in Section~\ref{sec:results} linear polarisation up to $52\pm6\%$ is detected. This high level of linear polarisation implies the presence of a highly ordered magnetic field in the emission region. In a simple model where the magnetic field consists of a uniform component with strength $B_0$ and a random component with strength $B_\text{rand}$, then the observed polarisation degree, $\Pi_\text{obs}$ is given by:

\begin{equation}
    \Pi_\text{obs} = \Pi_\text{max} \frac{B_0^2}{B_0^2 + B_\text{rand}^2} \zeta , 
\end{equation}

\noindent where $\zeta$ is some factor $<1$ taking into account other depolarisation effects \citep{burn_1966}. We find $\frac{B_0}{B_\text{rand}} \gtrsim 1.7$. This is consistent with a shock model, where shock compression of an initially turbulent magnetic field can leads to a more ordered field, in at least 1 dimension \citep{laing_1980}. The observed de-rotated polarisation angle (known as the electric vector position angle or EVPA), is not strictly parallel or perpendicular to the shock front as seen in Appendix~\ref{sec:pol_inten}, suggesting a more complex magnetic field structure across the shock front than a simple compressed slab, potentially indicating strong turbulence at the shock front.

The X-ray emission in the shock regions is co-spatial with the radio emission, but \cite{sell_2010} note that both a thermal and a synchrotron model can fit the X-ray spectrum equally well. They disfavour the thermal model due to requiring a low inclination in order to make the radio and X-ray emission appear co-spatial. In the thermal emission scenario the X-ray emission originates from the shocked ISM, expected to be pushed ahead of the expanding radio outflow. For example, this is clearly demonstrated in the case of the XRB S26 \citep{soria_2010}. \cite{sell_2010} further disfavour the thermal model due to the good correspondence of the measured power law index of the X-ray emission to the expectation, $\alpha=-1$, for a population of cooled synchrotron electrons with continuous injection of new ultra-relativistic electrons with $p\approx2$ \citep{kardashev_1962}.

We now have further evidence against a thermal origin for the X-ray emission. This is on the basis that if the thermal model holds, \cite{sell_2010} found a shock electron temperature of $6.6^{+2.2}_{-1.8}$~keV, which they used to calculate a shock velocity, $u_s$, of $2.4\times10^8~\text{cm s}^{-1}$ using:

\begin{equation}\label{eq:temp_vel_relation}
    kT = \frac{3}{16} \mu m_\text{P} u_s^2 ,
\end{equation}

\noindent where $\mu$ is the mean mass per particle ($\sim0.6$ for solar abundances) and $m_\text{P}$ is the proton mass. This assumes a strong shock (e.g. high Mach number), and thermal equilibrium between ion, proton and electron species in the shocked plasma at a temperature $T$. This temperature is too low to be consistent with our measured apparent velocity of $3\times10^9~\text{cm s}^{-1}$ in the standard shock heating scenario. Caveats to this argument which would result in a lower electron temperature for our known shock velocity are if the electrons and protons are not in thermal equilibrium, or if the shock is heavily cosmic ray dominated (e.g. \citealt{helder_2009}). In principle, one or a combination of these effects could explain the shock velocity and electron temperature observed as they both act to reduce the observed electron temperature for a given shock velocity. However, particle-in-cell simulations, which include both non-equilibrium effects and cosmic rays, suggest that in general the electron temperature is well predicted to within an order of magnitude by equation \eqref{eq:temp_vel_relation} (e.g. \citealt{gupta_2024}), making a large reduction in temperature from the mechanisms discussed above unlikely.

More evidence against the thermal model is found when we consider that the low inclinations for the shocks required in this case are incompatible with other aspects of our observations, as discussed further below. Therefore overall, we significantly favour a synchrotron origin for the radio and X-ray emission from these shocks. 


\subsection{Geometry of the powering outflow}\label{sec:geometry}

Given the emission mechanism of the shocks is known, we can now investigate the geometry of the powering outflow. We identify 3 distinct possibilities:

\begin{enumerate}
    \item A collimated straight jet
    \item A collimated precessing jet
    \item A wide-angle conical outflow
\end{enumerate}

In each of the 3 cases both a transient or continuously powered outflow is possible (see Section~\ref{sec:outflow_nature}), but the choice does not effect the analysis below. 

In the case of i), a well-collimated straight (constant launch direction) jet can produce an apparently wide shock structure in two scenarios. Either, the jet is inclined sufficiently close to the line of sight in so that the opening angle of the jet appears larger, known as foreshortening, or the wide shock structure is due instead to a jet heated cocoon/lobe of material driving a wider bow shock structure into the surrounding medium (\citealt{blandford_1974}, \citealt{kaiser_1997}, \citealt{kaiser_2004}). This is thought to be the case for the large scale shock structures seen in GRS 1915 \citep{motta_2025} and Cyg X-1 \citep{gallo_2005}. We strongly disfavour the latter of these scenarios due to the lack of significant curvature in the shock structure, as is seen in the other known examples. Furthermore, the hotspot, which in this cocoon model represents the location of the jet termination shock, is observed to vary in position angle between epochs for the shocks of Cir X-1. This is directly counter to the expectation for a jet with constant direction. If we do attempt to use this model to estimate the power of the jets (equation 10 in \citealt{motta_2025}), we obtain values of the jet power which are many orders of magnitude larger than the Eddington limit for a neutron star. This indicates that this model, where the wide shock we see is a bow shock which is driven by jet heated cocoon material, is not applicable in the case of Cir X-1. Instead, given the large shock velocity, the shocks are the direct site of interaction between the outflow from Cir X-1 and the ambient medium.


A collimated straight jet may still produce apparently wide-angled shocks if the inclination to the line of sight is sufficiently low. However, constraints derived from the apparent flux symmetry of the shocks, and their measured apparent velocities require the inclination angle to the line of sight to be $\gtrsim35\degree$, ruling out this scenario. In particular, the velocity of the shocks as a fraction of $c$, $\beta$, is given by the apparent velocity, $\beta_\text{app}$, and the inclination to the line of sight, $i$:

\begin{equation}\label{beta_from_app}
    \beta = \frac{\beta_\text{app}}{\sin i \pm \beta_\text{app} \cos i} , 
\end{equation}

\noindent where the $\pm$ refers to the approaching and receding shock respectively. We expect the bulk velocity of the emitting material to be approximately equal to the shock velocity in the strong shock regime. From our observations we can constrain the ratio of flux density between the two shocks to be $<3$, and by considering Doppler boosting, we can write:
\begin{equation}\label{eq:doppler_ratio}
    \left(\frac{1+\beta \cos i}{1-\beta \cos i}\right)^{3-\alpha} < 3 , 
\end{equation} 

\noindent where we take the spectral index of the emitting material, $\alpha$, to be $-0.5$. Solving this inequality by substituting in equation~\eqref{beta_from_app} and our measured value of the apparent shock velocity at a 2 pc projected distance (from Table~\ref{tab:position_fits}) we arrive at the conclusion that the inclination to the line of sight of the shocks and the powering outflow is $\gtrsim35\degree$.  We note this argument assumes that the intrinsic radio luminosity of each shock is the same. If this is not the case, due to an asymmetric outflow power, or asymmetric environment, then the situation can be fine tuned so that significant Doppler boosting is present (e.g. inequality \ref{eq:doppler_ratio} does not hold), but is not observed as the approaching material is intrinsically fainter. This scenario is unlikely, and in this case would require significant asymmetry in the shock luminosity to invalidate our inclination constraint above.

We also obtain a less constraining, but robust limit on the inclination angle of $>9\degree$ from considering where $\beta>1$ for the receding case of equation~\eqref{beta_from_app}. Given these constraints on inclination, and the measured apparent velocities in Table~\ref{tab:position_fits}, we conservatively take the de-projected speed of the shocks at a propgation distance of 2~pc to be $\approx0.1c$ for calculations in Sections~\ref{sec:energetics} and \ref{sec:particle_accel}. Projection effects could lead to the true de-projected speed being up to a factor of two greater than this.

The true half opening angle of the shocks, $\psi$, is then given by the observed opening angle $\psi_\text{obs}$ and the inclination angle:

\begin{equation} \label{eq:opening_angle_inc}
      \tan \psi = \Gamma\tan \psi_\text{obs}  \sin i , 
\end{equation}

\noindent where the factor of $\Gamma$, the bulk Lorentz factor, comes from considering time dilation effects \citep{miller-jones_2006}, but can be neglected in our case as $\Gamma\sim1$. The limit on the inclination angle, combined with the observed half opening angle of $\sim28\degree$, constrains the true half opening angle to be $>17\degree$. Therefore, these shocks cannot be powered by a collimated XRB jet which have half opening angles $\lesssim5\degree$ \citep{miller-jones_2006} unless these jets change direction over time.

It is then natural to consider case ii), a collimated jet which changes direction, through precession or some other mechanism, as a cause of these wide opening angle shocks. Cir X-1 is known to launch a collimated precessing relativistic jet \citep{cowie_2025}. However, while past jet observations of Cir X-1 have shown a jet angle consistent with the direction of the shocks (e.g. \citealt{tudose_2008}, \citealt{miller-jones_2011b}), other observations (from late 2011 onwards) have shown a jet position angle inconsistent with the shocks (\citealt{coriat_2019}, \citealt{cowie_2025}). This implies at the very least that the precessing jet of Cir X-1 is not continuously powering these shocks, raising powering requirements, which are discussed further in Section~\ref{sec:energetics}. Furthermore, the absence of similar shock structures in the current direction of the precessing jet seems to raise the possibility that the known precessing jets in this source are not responsible for powering the shocks, unless some recent (e.g. 2011 onwards) mode change in the precessing jet power has occurred. Therefore, we disfavour a precessing jet powering mechanism for the shocks, but given the complexity of the jet behaviour observed from this source (see \citealt{cowie_2025} for an overview), it cannot be ruled out.

The only remaining scenario is iii), a truly wide relativistic outflow launched from Cir X-1. Given the inferred launch velocities of the shocks from Table~\ref{tab:position_fits}, and the fact that the outflow velocity is at least that of the shocks \citep{marti_1997}, this outflow must have $\beta\gtrsim0.3$ and an opening angle $>17\degree$. An outflow with such properties is unprecedented in XRBs, perhaps to be expected given the youth and peculiarity of Cir X-1. In order to better understand this outflow we can use the shocks where it interacts with the ambient medium to measure the power of this outflow.

\subsection{Energetics of the shocks}\label{sec:energetics}

We can estimate the energy and other physical parameters of the shocks using the observed radio and X-ray synchrotron emission, their geometric properties, and their velocity. 

\subsubsection{Internal energy from equipartition}\label{sec:energy_internal}

It can be shown for a homogeneous synchrotron emitting plasma of known volume, that the energy required to produce a given synchrotron luminosity is a function of the magnetic field, and has a minimum close to equipartition of energy between the magnetic field and radiating particle population (\citealt{burbidge_1956}, \citealt{pacholczyk}). It is assumed that the particle population is non-thermal, and described by a power law with a low and high energy cut-off. This leads to a low and high frequency cut-off in the synchrotron spectrum ($\nu_\text{min}$ and $\nu_\text{max}$). The equations for the required minimum energy and associated magnetic field strength, as a function of observable parameters are:

\begin{align} \label{eq:equip_energy}
    E_\text{min} & = \frac{7}{4}(1+\eta)^{\frac{4}{7}} \left(\frac{6 \pi}{fV} \right)^{-\frac{3}{7}} \nonumber  \\ & \cdot \left( \frac{2 c_2^{-1} c_1^{\frac{1}{2}}}{1+2\alpha}   4\pi D^2 F_{\nu_\text{obs}} \nu_\text{obs}^{-\alpha}\left(\nu_\text{max}^\frac{1+2\alpha}{2}-\nu_\text{min}^\frac{1+2\alpha}{2}\right) \right)^\frac{4}{7} ,
\end{align}

\begin{align} \label{eq:equip_B}
    B_\text{eq}  &  = \left(\frac{6 \pi (1+\eta)}{fV} \frac{2 c_2^{-1} c_1^{\frac{1}{2}}}{1+2\alpha}   4\pi D^2 F_{\nu_\text{obs}} \nu_\text{obs}^{-\alpha}\left(\nu_\text{max}^\frac{1+2\alpha}{2}-\nu_\text{min}^\frac{1+2\alpha}{2}\right)\right)^\frac{2}{7} , 
\end{align}

\noindent where  $V$ is the observed volume of emission and $f$ is the filling factor of the plasma in this volume. $D$ is the distance to the source, and $F_{\nu_\text{obs}}$ is the flux density of the synchrotron emission at some observation frequency $\nu_\text{obs}$. $\alpha$ is the spectral index of the synchrotron spectrum, $\eta$ is the fraction of energy in non-thermal protons relative to non-thermal electrons (assumed to be 0 for a minimum energy estimate), and $c_1$, $c_2$ are constants. A full derivation of these formulae, along with definitions of the constants can be found in \cite{pacholczyk} and is also provided in Appendix~\ref{sec:energetics_appen}. The above equations and those throughout this section do not include potential corrections for Doppler boosting due to bulk relativistic motion of the emitting material. As discussed in Section~\ref{sec:geometry} this is unlikely to be a significant effect for the shocks of Cir X-1 so can be neglected. Once we have an energy in non-thermal electrons, $E_e$, and a magnetic field measurement, $B$, regardless of whether these are the equipartition estimates, a total number of non-thermal electrons can be obtained:

\begin{equation} \label{eq:electron_number}
    N_e = E_e \frac{1+2\alpha}{2\alpha} c_1^\frac{3}{2}B_\text{eq}^\frac{3}{2} \frac{\left(\nu_\text{max}^\alpha - \nu_\text{min}^\alpha\right)}{\left(\nu_\text{max}^\frac{1+2\alpha}{2} - \nu_\text{min}^\frac{1+2\alpha}{2}\right)}
\end{equation}

\noindent In order to calculate the minimum energy of the shocks and related parameters we assume an ellipsoidal geometry for the shocks with semi-axes in the plane of the sky of 1.3 and 0.9 pc, corresponding to the region used in Section~\ref{sec:results} to measure the flux density, and the assumption that the semi-axis along the line of sight is 1 pc. This results in a volume of $4.4\pm0.8\:\text{pc}^3$, where the uncertainty is dominated by the uncertainty on the known distance of $9.4\pm1$~kpc \citep{heinz_2015}. We assume a filling factor of 1, and discuss this assumption later in this section. We use our measured values of flux density at $2.6$~GHz for the NW shock, $8\pm2$~mJy, and a spectral index of $-0.6\pm0.2$, where the large uncertainty reflects the different values of $\alpha$ which have been measured for the shocks (\citealt{sell_2010}, \textcolor{blue}{Gasealahwe, Savard et al. 2025}). \cite{sell_2010} observed an X-ray spectral index value indicating the X-ray emission at $\sim\text{keV}$ energies is above the synchrotron cooling break. They estimated the synchrotron cooling break to occur at a frequency $\nu_\text{cool}=2\times10^{16}$~Hz by extrapolating the radio and X-ray spectra. This indicates that the energy spectrum of particles is not well modelled by a single power law for electrons which radiate at frequencies above $\nu_\text{cool}$. For these estimates we conservatively take $\nu_\text{max}$ to be equal to $\nu_\text{cool}$, effectively disregarding the energy contribution from electrons radiating above the cooling break. A more stringent treatment (see Section~\ref{sec:particle_accel} for an example) would fully model the particle injection and cooling. Our treatment here leads to a slight underestimate (by a few \%) of $E_\text{min}$ and $B_\text{eq}$ given the weak dependence on $\nu_\text{max}$. Finally, $\nu_\text{min}$ is not measured in any of our observations, but can be constrained to be between $1$~GHz, as we do not observe the frequency break to a $\alpha=\frac{1}{3}$ expected below $\nu_\text{min}$, and $10$~kHz. The lower bound is a result of the magnetic field strength of $\sim100\:\mu\text{G}$ \citep{sell_2010}, leading to $\nu_\text{cr}$ for a synchrotron emitting plasma being always greater than $10$~kHz (see equation \ref{eq:crit_freq}).

Given the large uncertainties on some of the parameters required to calculate $E_\text{min}$ and $B_\text{eq}$, and the various strong and weak dependencies in equations \eqref{eq:equip_energy} to \eqref{eq:electron_number}, we adopt a Monte Carlo sampling approach to understand the uncertainties on our calculated values of $E_\text{min}$, $B_\text{eq}$ and $N_e$. Overall we find that the sampled distributions for each of our calculated parameters do not appear to be well described by standard analytic forms, but are clearly unimodal. Therefore, below we report the empirical confidence interval defined as the range spanned by the central 68\% of samples. This corresponds to the commonly used Gaussian-equivalent $1\sigma$. The full sampled distributions and the 68\%, 95\%, and 99.7\% confidence intervals are presented in Appendix~\ref{sec:energetics_appen} for completeness.

We find the minimum energy required to create the synchrotron emission observed from one of the shocks is $
\log_{10}(E_\text{min} [\text{ergs}])=46.1^{+0.3}_{-0.2}$ and the magnetic field corresponding to this minimum energy (equipartition) case to be $\log_{10}(B_\text{eq}[\text{G}])=-4.53^{+0.16}_{-0.12}$. These measurements are similar to the estimates of $8\times10^{45}\:\text{erg}$ and $\sim 50\:\mu\text{G}$ made in \cite{sell_2010}. Finally, we find the number of non-thermal electrons in the equipartition case to be $\log_{10}(N_e)= 48.7^{+1.2}_{-0.9}$. Using the most conservative (longest) estimate of the lifetime of the shocks from the constant deceleration model of $50\pm5$ years, we therefore obtain an average power required simply to create the synchrotron emitting particles and magnetic field of $\log_{10}(Q_\text{min}[\text{erg s}^{-1}])=36.9^{+0.3}_{-0.2}$. We note this is almost two orders of magnitude greater than the estimate in \cite{sell_2010}. They used the extrapolated cooling break frequency and equipartition magnetic field to derive the synchrotron cooling timescale, which they took to be the timescale of energy delivery to the shocks.

\subsubsection{Magnetic field estimate from cooling break} \label{sec:mag_field_from_cooling}

Given we now know the lifetime of the shocks from measurements of their proper motion and have a constraint on the break frequency $\nu_\text{cool} \sim 2\times10^{16}~\text{Hz}$ from \cite{sell_2010}, we can obtain an independent estimate of the magnetic field, without assuming equipartition. We can find the required magnetic field to cause significant synchrotron cooling of electrons radiating at the break frequency over the lifetime of the shocks using the equation for the synchrotron cooling timescale:

\begin{equation}
    t_\text{sync} = \frac{6\pi m_e^2 c^3 c_1^\frac{1}{2}}{ \sigma_t (B \sin \theta) ^\frac{3}{2} \nu_\text{cool}^\frac{1}{2}} ,
\end{equation}
 
\noindent where $\sigma_t$ is the Thomson cross section and $\theta$ is the angle between the magnetic field and the electron velocity, where a value of $\frac{\pi}{2}$ is assumed hereafter. Rearranging for $B$, we find given $t_\text{sync} < 50~\text{years}$ and $\nu_\text{cool} < 1\times10^{17}~\text{Hz}$, that $B>200~\mu\text{G}$. This value for the magnetic field strength is in tension at the $\sim5\sigma$ level with the equipartition field strength estimated from observations. This tension could arise in a few scenarios.

First, we note that if another cooling mechanism contributes significantly, namely inverse Compton losses, the synchrotron cooling time could be longer than the lifetime of the shocks and the frequency break could be due to inverse Compton cooling, removing the tension in the magnetic field estimates. However, as will be shown in Section~\ref{sec:particle_accel}, inverse Compton, and indeed all other cooling mechanisms, are sub-dominant to synchrotron cooling for electrons radiating above the cooling break. 

Second, the equipartition estimate of the magnetic field could be raised to be in line with the constraint from the cooling break. This can happen if either the filling factor is decreased, or non-thermal protons introduced which do not contribute to the observed synchrotron luminosity, but allow for equipartition between non-thermal particles and the magnetic field at larger magnetic field values. To raise the equipartition estimate of the magnetic field to $>200~\mu\text{G}$ the filling factor would need to be reduced to $\sim10^{-3}$. This is unlikely given that a relativistic plasma expanding at a fraction of the speed of light ($\sim0.1c$) would for a scale size of $\sim1~\text{pc}$, expand to greater than this filling factor in $<1~\text{year}$, much shorter than the lifetime of the shock. To acheive the same effect with the introduction of non-thermal protons these protons would need to have a total energy $\eta\sim300$ times that in the non-thermal electrons. This fraction of energy in non-thermal protons is slightly larger than the $\eta\sim10-100$ expected from some observations and particle-in-cell simulations (e.g. \citealt{morlino_2012} \citealt{gupta_2024}), but is considered plausible. We note that $\eta\sim300$ increases the minimum energy and minimum power estimates by over an order of magnitude. 

Finally, equipartition may not hold for the system especially given the fairly young age of the shocks and the associated synchrotron plasma. This means the equipartition estimate for the magnetic field may not be correct. However, the deviation from equipartition needed for a magnetic field value consistent with the cooling estimate would increase the energy in the synchrotron plasma by an order of magnitude. Overall, it is likely that a combination of several of these factors: slightly decreased filling factor, presence of non-thermal protons, and a deviation from equipartition, are responsible for the tension between the equipartition and synchrotron cooling estimates of the magnetic field. However, these changes will overall result in a net increase to the calculated minimum energy in non-thermal particles and the magnetic field, and therefore the power required to create the shock structures.

\subsubsection{Kinetic energy from swept up mass}\label{sec:kinetic_energy}

We have now shown that the time-averaged power required to supply only the non-thermal particle and magnetic field energy of both shocks has a robust lower limit of $\sim10\%$ of the Eddington luminosity of a $1.4~\text{M}_\odot$ neutron star ($1.8\times10^{38}~\text{erg s}^{-1}$). We can also estimate the kinetic energy in the shocked volume, by calculating the mass swept up by the shocks, and the kinetic energy of this swept up material given the known shock speed. The volume swept up by the shocks depends on the model of the powering outflow, e.g. whether it is a wide conical outflow, or a precessing jet. As discussed in Section~\ref{sec:geometry} a conical outflow is more likely, but a correction factor to the volume can be applied for the precessing jet case. The volume swept out by a conical outflow is:

\begin{equation}\label{eq:swept_out_volume}
    V=\frac{1}{3} \pi h^3 \tan^2\psi = \frac{1}{3}\pi h_\text{obs}^{3} \tan^2{\psi_\text{obs}} (\sin{i})^{-1} ,
\end{equation}
\noindent where $h_\text{obs}$ is the projected distance of the shocks from Cir X-1, and the inclination dependence comes in part from equation~\eqref{eq:opening_angle_inc}. Taking the point in time when $h_\text{obs}=2~\text{pc}$ (see Table~\ref{tab:position_fits}) we obtain $V=2.8(\sin{i})^{-1}~\text{pc}^{3}$. For the case of a precessing jet with a jet opening angle, $\phi$, of $1-5\degree$ as is typical for XRB jets \citep{miller-jones_2006}, the volume swept out is a hollow cone, and equation~\ref{eq:swept_out_volume} is adjusted by a factor of:

\begin{equation}
    \frac{\tan^2{\psi_\text{obs}} - \tan^2{(\psi_\text{obs} - \phi)} }{\tan^2{\psi_\text{obs}}} \sim 0.08~\text{to}~0.35~\text{for}~\phi=1\degree~\text{to}~5\degree .
\end{equation} 


\noindent We also require an estimate of the density of the ambient material, $\rho_\text{amb}$, which is swept up by the shocks. Given the inclination constraints on the shocks, they are still propagating within the SNR environment of Cir X-1. Spectroscopy of the SNR X-ray emission showed that the ISM number density outside the SNR is $\sim0.3~\text{cm}^{-3}$ \citep{heinz_2013}. We can then turn to SNR evolution theory (\citealt{chevalier_1974}, \citealt{straka_1974}, \citealt{band_1988}), simulations of SNR (\citealt{villagran_2020}, \citealt{meyer_2024}), and specific simulations of the \emph{Africa nebula}, which include the launch of a powerful collimated jet at an early time in the SNR evolution (\textcolor{blue}{Gasealahwe, Savard et al. 2025}). All of these suggest that a value for the density near the centre of the SNR of $10^{-2}~\text{cm}^{-3}$ is conservatively low, given the known parameters of the SNR \citep{heinz_2013}. We may also consider the density profile of the SNR. This is not known for Cir X-1 due to the faintness of the X-ray emission from the SNR. However, its surface brightness in X-rays appears flat or slightly centrally peaked, in contrast with the most common morphology of SNRs where the X-ray emission has a shell morphology (e.g. \citealt{acero_2007}). This suggests that the \emph{Africa nebula} belongs to the mixed morphology class of SNRs (\citealt{long_1991}, \citealt{jones_1993}, \citealt{rho_1998}, \citealt{chiotellis_2024}). Mixed morphology SNRs are thought to be formed when the SNR interacts with a multi-phase ISM, in particular causing gas evaporation from clouds which survive passage through the SNR shock (\citealt{cowie_1981}, \citealt{white_1991}). These mixed-morphology remnants are notable in this case because simulations and models suggest a higher central density than expected from typical SNR dynamics, with densities within the SNR of $\sim1~\text{cm}^{-3}$, and a flat density profile. Given the uncertainty in the density profile within the SNR, when estimating the mass swept up by the shocks of Cir X-1 we assume a flat density profile using the conservative value of $10^{-2}~\text{cm}^{-3}$, expected from typical SNR dynamics, while noting that values of up to $\sim1~\text{cm}^{-3}$ may be possible.



For the volume above and number density of $10^{-2}~\text{cm}^{-3}$, assuming a hydrogen composition, we calculate a total number of swept up protons as $8\times10^{53}$ and a swept up mass of $1\times10^{30}~\text{g}$. Given that at $2~\text{pc}$ the apparent shock velocity, is $0.14c\pm0.03c$ and at smaller distances was greater than this due to the deceleration. Furthermore, the true de-projected velocity will be close to or larger than this (see equation~\ref{beta_from_app} and Section~\ref{sec:geometry}). Therefore,  we can conservatively estimate that the bulk velocity of the shocked material is $\sim0.1c$. The kinetic energy of the shocked material can then be calculated:

\begin{equation}
    E_k = \frac{1}{2}\rho_\text{amb}Vu_s^2 = \frac{\pi \rho_\text{amb}}{6} h_\text{obs}^3 \tan^2 \psi_\text{obs} u_{s,\text{obs}}^2 (\sin i)^{-3} ,
\end{equation}

\noindent resulting in a total kinetic energy budget for one of the shocks of \textbf{at least $\mathbf{5\times10^{48}}~\text{erg}$}. Over a lifetime of 50 years, this requires a time-averaged power of:

\begin{equation}
    P_\text{kin} \approx 4\times10^{39}\left(\frac{n}{10^{-2} \text{cm}^{-3}}\right)~\text{erg s}^{-1} = \mathbf{20L_\textbf{Edd}} \cdot \left(\frac{n}{10^{-2} \text{cm}^{-3}}\right) 
\end{equation}

This order of magnitude estimate for the kinetic energy of the shocks reveals the extreme power required to create these structures, and that the energy output of this accreting neutron star is apparently dominated by the wide-angle mildly relativistic powering outflow.

\subsection{Particle acceleration properties}\label{sec:particle_accel}

Measuring the shock velocity and other properties not only allows us to understand the macroscopic properties but also the details of the particle acceleration occurring at the shocks, which creates the non-thermal particle population responsible for the observed synchrotron radiation (see \citealt{matthews_2020} and references therein for a review). 

\subsubsection{Maximum energy of accelerated particles}

We begin by understanding the maximum energy which particles can be accelerated to in the shocks. The confinement energy is given by:

\begin{equation}
    E_\text{conf} = ZeBR , 
\end{equation}

\noindent where $Z$ is the particle charge in units of the electron charge, $e$, and $R$ is the scale size of the particle acceleration region, taken to be the size of the downstream region of the shock, in our case $\sim1~\text{pc}$. The shocks of Cir X-1 have a confinement energy, for protons and electrons, of $\sim200~\text{PeV}$. We can then consider the Hillas energy \citep{hillas_1984}, which can be interpreted, among other ways, as the timescale balance between the dynamical time of the system and the acceleration time, assuming the Bohm diffusion limit, of a particle undergoing diffusive shock acceleration (DSA) \citep{bell_1978}. The dynamic timescale is:

\begin{equation}\label{eq:esc_time}
    t_\text{dyn} = \frac{R}{u_s} ,
\end{equation}

\noindent where $u_s$ is the shock velocity. The acceleration timescale of DSA, with a diffusion coefficient of $D=\frac{\kappa r_g c}{3}$, where $r_g$ is the particle gyro-radius, is given by:

\begin{equation}\label{eq:dsa_accel}
\begin{split}
    t_\text{acc} &= \frac{D}{u_s^2} = \frac{\kappa r_g c}{3u_s^2} = \frac{\kappa E c}{3ZeBu_s^2} \\&= 4\times10^7 \frac{\kappa}{Z}\left(\frac{E}{1\:\text{PeV}}\right) \left(\frac{B}{1\:\mu\text{G}}\right)^{-1} \left(\frac{u_s}{c}\right)^{-2} \:\text{s}.
\end{split}
\end{equation}

\noindent Then equating equations~\eqref{eq:esc_time} and \eqref{eq:dsa_accel}, and taking the Bohm diffusion limit, $\kappa=1$ we find the Hillas energy:

\begin{equation}
    E_H = 10^{15} \text{eV} \left(\frac{R}{1\: \text{pc}}\right) \left(\frac{B}{1\: \mu\text{G}}\right) \left(\frac{u_s}{c}\right) Z .
\end{equation}

\noindent For the shocks of Cir X-1, the shock velocity is at least $0.1c$ (Section~\ref{sec:geometry}), the magnetic field is $>200~\mu\text{G}$ (Section~\ref{sec:mag_field_from_cooling}) and the scale size of the acceleration region is taken to be the size of the downstream region of the shock, $\sim1~\text{pc}$. This gives a Hillas energy of $\sim20~\text{PeV}$, noting that this energy limit also applies for acceleration mechanisms other than DSA. We can also examine the constraint on the maximum particle energy from the power provided by the central source, Cir X-1. The Hillas-Waxman-Lovelace power limit \citep{lovelace_1976} is:


\begin{equation}
    Q_{k,\text{min}} = 10^{35} \: \text{erg s}^{-1} ~ \epsilon^{-1}\left(\frac{E}{1 \: \text{PeV}}\right)^2 \left(\frac{u_s}{c}\right)^{-1} Z^{-2} , 
\end{equation}

\noindent where $\epsilon$ is a conversion efficiency of kinetic to magnetic energy, and for DSA can be thought of as an efficiency of magnetic field amplification at the shock, and a value of $\sim0.1$ is often assumed \citep{matthews_2020}. This leads to a required power of $\sim10^{39}~\text{erg s}^{-1}$, which is close to the estimated total power of the outflow in Section~\ref{sec:kinetic_energy}. This indicates that the power of the outflow may also be a limiting factor for particle acceleration alongside the Hillas energy limit. 

Finally, we can consider in more detail the loss versus acceleration timescales for the case where the particles are being accelerated through DSA, which seems likely given the shock nature and polarisation properties. In particular, the detection of high linear polarisation fractions, implies the ordered and turbulent magnetic fields are of roughly the same strength, which is often referred to as the strong turbulence limit. This magnetised turbulence may be driven by cosmic rays and is thought to be a necessary condition for particle acceleration close to or at the Bohm limit (e.g. $\kappa=1$) \citep{bell_2004}.  

For electrons, we examine the different loss timescales to determine which one is dominant. To do this we use the \textsc{gamera} package \citep{gamera} in Python to determine the loss timescales for a population of electrons in a $200~\mu\text{G}$ magnetic field, with an ambient medium density of $10^{-2}~\text{cm}^{-3}$, and a background radiation field made up of the CMB, the interstellar radiation field (using the model presented in \citealt{popescu_2017}), and the diluted radiation field from the host star. We assume the host star to be a typical high mass O-type star ($10~R_\odot$, $4\times10^{4}~\text{K}$ blackbody), in order to maximise the radiation field, and hence losses due to inverse Compton, at a distance of $\sim2~\text{pc}$. Finally, we also add adiabatic cooling of the electron population, assuming a 1~pc region expanding at $0.1c$ as a maximal case. We find, for the high energy particles, $>10~\text{TeV}$ synchrotron losses dominate over all other loss mechanisms by over an order of magnitude.

The loss timescale for synchrotron cooling is given by:

\begin{equation}
\begin{split}
    t_\text{sync} = \frac{E}{\dot{E}_\text{sync}} = \frac{3m_ec^2\gamma_e8\pi}{4\sigma_T c \gamma_e^2 B^2} = 4\times10^{11} \left(\frac{E}{1\:\text{PeV}}\right)^{-1} \left(\frac{B}{1\:\mu\text{G}}\right)^{-2} \: \text{s} , 
\end{split}
\end{equation}

\noindent where $\gamma_e$ is the Lorentz factor of the electron. Equating this loss timescale with the acceleration timescale for DSA from equation~\eqref{eq:dsa_accel} we obtain a maximum energy achievable by electrons for a $0.1c$ shock velocity, of $\sim0.7~\text{PeV}$. Given this predicted maximum energy to which electrons can be accelerated we can predict an exponential cut-off in the the synchrotron spectrum above $\sim7~\text{MeV}$. This maximum predicted energy requires the electrons to be undergoing particle acceleration in the Bohm limit. The observation of X-ray synchrotron radiation up to a photon energy of $10~\text{keV}$ requires electrons with an energy of least $30~\text{TeV}$ (in a $200~\mu\text{G}$ magnetic field, using equation~\ref{eq:crit_freq}). This is direct evidence of particle acceleration up to at least $\sim5\%$ of the maximum energy in the idealised Bohm limit. We note that this ratio between the maximum predicted energy in the Bohm limit and the observed energy of the electrons is independent of the assumed magnetic field.

We can also consider the loss timescales for the proton case, although inverse Compton and synchrotron losses are sub-dominant due to the mass dependence of these mechanisms. Therefore, we instead consider losses from inelastic collisions with thermal protons, photo-meson production and pair production \citep{begelman_1990}. We find for each of these cases, by comparison to the acceleration timescale, that the cooling timescales do not limit particle acceleration to the PeV regime. This is expected for thermal proton collisions \citep{begelman_1990}, and for the photo-meson and pair production the cooling time is long given the relatively low energy density of the photon background. Therefore, the limiting factor on the maximum energy protons can achieve in the shocks is set by the Hillas condition. This makes these shocks, powered by an accreting neutron star, \textbf{potential sites of acceleration of protons up to} $\mathbf{\sim20~}\textbf{PeV}$. The population of XRBs have been postulated to potentially explain cosmic rays above the "knee" (\citealt{heinz_2002b}, \citealt{fender_2005}, \citealt{cooper_2020}, \citealt{wang_2025}), and our analysis of this case shows that certain large scale outflow interaction structures may provide the necessary conditions for this particle acceleration to high energies to occur.

Given our derived maximum energy cut-off and  the assumption that the energy in non-thermal protons is $\eta\sim100$ times the energy in the non-thermal electrons, we can use equation~\eqref{eq:equip_energy} to estimate the total energy in cosmic rays (non-thermal protons). We find $E_\text{min}\sim1.8\times10^{47}~\text{erg}$ of which $1\times10^{47}~\text{erg}$ is in cosmic rays (the rest of the energy being in the magnetic field, see Appendix~\ref{sec:energetics_appen}). Using the $\sim50~\text{year}$ measured lifetime of the shocks this gives a cosmic ray luminosity by the shocks of $\sim6\times10^{37}~\text{erg s}^{-1}
$. Then, assuming a power law cosmic ray spectrum with energy index of $-2$, a low energy cut-off of the proton rest mass, and a high energy cut-off of $20~\text{PeV}$, we can estimate the cosmic ray luminosity of Cir X-1 at energies between $10~\text{TeV}$ and $20~\text{PeV}$, e.g. around the "knee". We find a cosmic ray luminosity in this energy range of $\sim5\times10^{36}~\text{erg s}^{-1}$. This is not particularly sensitive to the exact values of the energy index or cut-offs. The total cosmic ray luminosity at Earth at these energies can be calculated assuming a total luminosity of $\sim1\times10^{41}~\text{erg s}^{-1}$, an energy index of $-2.7$, and a low energy cut-off of the proton rest mass (see \citealt{grenier_2015} and references therein). The choice of high energy cut-off has a negligible effect. The total cosmic ray luminosity at Earth between $10~\text{TeV}$ and $20~\text{PeV}$ is then $\sim2\times10^{38}~\text{erg s}^{-1}$. Therefore, with reasonable assumptions on the fraction of energy contained within non-thermal, the shocks of Cir X-1 could contribute a few percent of the observed cosmic ray luminosity at energies around the "knee". The cosmic rays from the shocks of Cir X-1 will not have reached Earth yet, and the measured cosmic ray luminosity is averaged on a much longer timescale than the lifetime of the shocks due to diffusion of cosmic rays. However, considering there are several powerful active X-ray binaries in our galaxy at any given time, these sources, in particular those with large scale structures, are possibly a dominant source of cosmic rays around the "knee". This has been suggested by both theoretical studies e.g. \cite{cooper_2020} and possibly by recent observational results e.g. \cite{lhaaso_2024}.

\subsubsection{Efficiency of particle acceleration}


As well as the maximum energy of accelerated particles, we can use our estimates of the energy in non-thermal electrons and our estimates of the shock kinetic energy to attempt to calculate the injection and energetic efficiencies of the electron acceleration mechanism. We define these as: the proportion of the cold electron population swept up by the shock which is accelerated to become non-thermal, and the proportion of the shock bulk kinetic energy which is in non-thermal electrons, respectively: 

\begin{equation}
    \epsilon_\text{inj} = \frac{N_{e,\text{nt}}}{N_{e,\text{sw}}}
\end{equation}

\begin{equation}
    \epsilon_e = \frac{E_e}{E_\text{kin}} = \frac{E_e}{\frac{1}{2}m_\text{P} u_s^2 N_{e,\text{sw}}}, 
\end{equation}

\noindent where $N_{e,\text{sw}}$ and $N_{e,\text{nt}}$ are the total number of swept up electrons and non-thermal electrons respectively, $E_e$, is the total energy in non-thermal electrons, and $E_\text{kin}$ is the total bulk kinetic energy of the shock (e.g. the bulk kinetic energy density of the initial upstream flow in the shock frame), as calculated in Section~\ref{sec:kinetic_energy}. This assumes an ionised hydrogen composition to the medium swept up by the shock. 

In order to calculate the efficiencies for the shock we assume the particle acceleration is dominated by a forward shock propagating into the ambient medium. This may not be the case if instead a reverse shock propagating into the outflow material is responsible for accelerating the majority of non-thermal particles which produce the observed synchrotron radiation. 

Our observations reveal only the presence of a single sharp drop in surface brightness on the far side of the cap structure from Cir X-1 (see Figure~\ref{fig:flux_cut}). This could be taken to be either the site of the forward shock, or the contact discontinuity. This depends on whether the forward shock accelerates particles which produce synchrotron emission. Other sharp changes in the surface brightness profile may be hidden by back-flow or other projected structures which give rise to the slow decay of the surface brightness profile towards Cir X-1 (see Figure~\ref{fig:flux_cut}). 

If the reverse shock is responsible for particle acceleration the sharp drop in surface brightness we observe with proper motion is then the contact discontinuity. The number of particles encountering the relevant shock front (in this case the reverse shock) is then more uncertain, as we do not know the density of the outflow through which it propagates. However, we can somewhat constrain the density of the outflow material from momentum balance at this working surface (contact discontinuity) \citep{marti_1997}. We find that to produce the mildly relativistic advance speed of the working surface, the outflow material cannot be significantly more undersense than the ambient medium without requiring highly relativistic outflow velocities. Therefore, while the efficiencies calculated in this Section are nominally for the case where the forward shock is responsible for the majority of the particle acceleration, they are likely upper limits in the reverse shock case given their scaling with the density of the medium crossing the shock.

We emphasise, that outside of the efficiencies calculated in this Section, all other physical quantities estimated, such as energetics, are agnostic to whether a reverse or forward shock is responsible for the majority of particle acceleration. For example, the swept up mass calculation in Section~\ref{sec:kinetic_energy} applies to the forward shock regardless of whether it is responsible for the majority of particle acceleration.

Beginning with $\epsilon_e$, we use that the total energy in non-thermal electrons, $E_e$, is related to our minimum internal energy estimate by $E_e = \frac{4E_\text{min}}{7(\eta+1)}$, where $\eta$ is the ratio of energy in non-thermal protons to non-thermal electrons. Then $E_e\propto(\eta+1)^{-\frac{3}{7}}$, as $E_\text{min}\propto(\eta+1)^{\frac{4}{7}}$. We can then estimate the energetic electron acceleration efficiency to be:

\begin{equation}
 \epsilon_e \approx 3\times10^{-3} \cdot (1+\eta)^{-\frac{3}{7}} \cdot \left(\frac{n}{10^{-2} \text{cm}^{-3}}\right)^{-1} ,
\end{equation}




\noindent where the scaling with average ambient number density, $n$, comes from the shock bulk kinetic energy calculated in Section~\ref{sec:kinetic_energy}. 

The electron injection efficiency estimate is more uncertain given our order of magnitude uncertainties on the total number of non-thermal electrons, largely driven by the uncertainty on the low energy cut-off to the electron energy spectrum (see Section~\ref{sec:energy_internal}). Given we estimated the total number of swept up ambient protons (and hence electrons) to be $8\times10^{53}$ in Section~\ref{sec:kinetic_energy}, then we can estimate the electron injection efficiency as:

\begin{equation}
    \epsilon_\text{inj} \sim 6\times10^{-6} \cdot (1+\eta)^{-\frac{3}{7}} \cdot \left(\frac{n}{10^{-2} \text{cm}^{-3}}\right)^{-1} ,
\end{equation}

\noindent with approximately an order of magnitude uncertainty. The scaling of $N_{e,\text{nt}}$ with $\eta$ is the same as that of $E_e$. This means that $\epsilon_e$ and $\epsilon_\text{inj}$ have the same scaling with $\eta$ and $n$, allowing for the ratio to be estimated as:

\begin{equation}
    \frac{\epsilon_\text{inj}}{\epsilon_e} \approx 2\times10^{-3} , 
\end{equation}

\noindent with approximately an order of magnitude uncertainty.

The shocks of Cir X-1 are now one of a handful of astrophysical systems where these efficiencies have been estimated, paving the way for better understanding of the physics of shocks and particle acceleration. Other systems include S26, where an electron energetic efficiency of $\sim10^{-3}$ was inferred \citep{soria_2010}, or the SNR Tycho, where an electron energetic efficiency of $\sim10^{-4}$ was observed \citep{morlino_2012}. These observationally derived efficiencies are consonant with numerical particle-in-cell experiments (e.g. \citealt{sironi_2011}, \citealt{park_2015}, \citealt{gupta_2024}).


\subsection{SED modelling}\label{sec:sed}

\begin{figure*}
 \includegraphics[width=1.9\columnwidth]{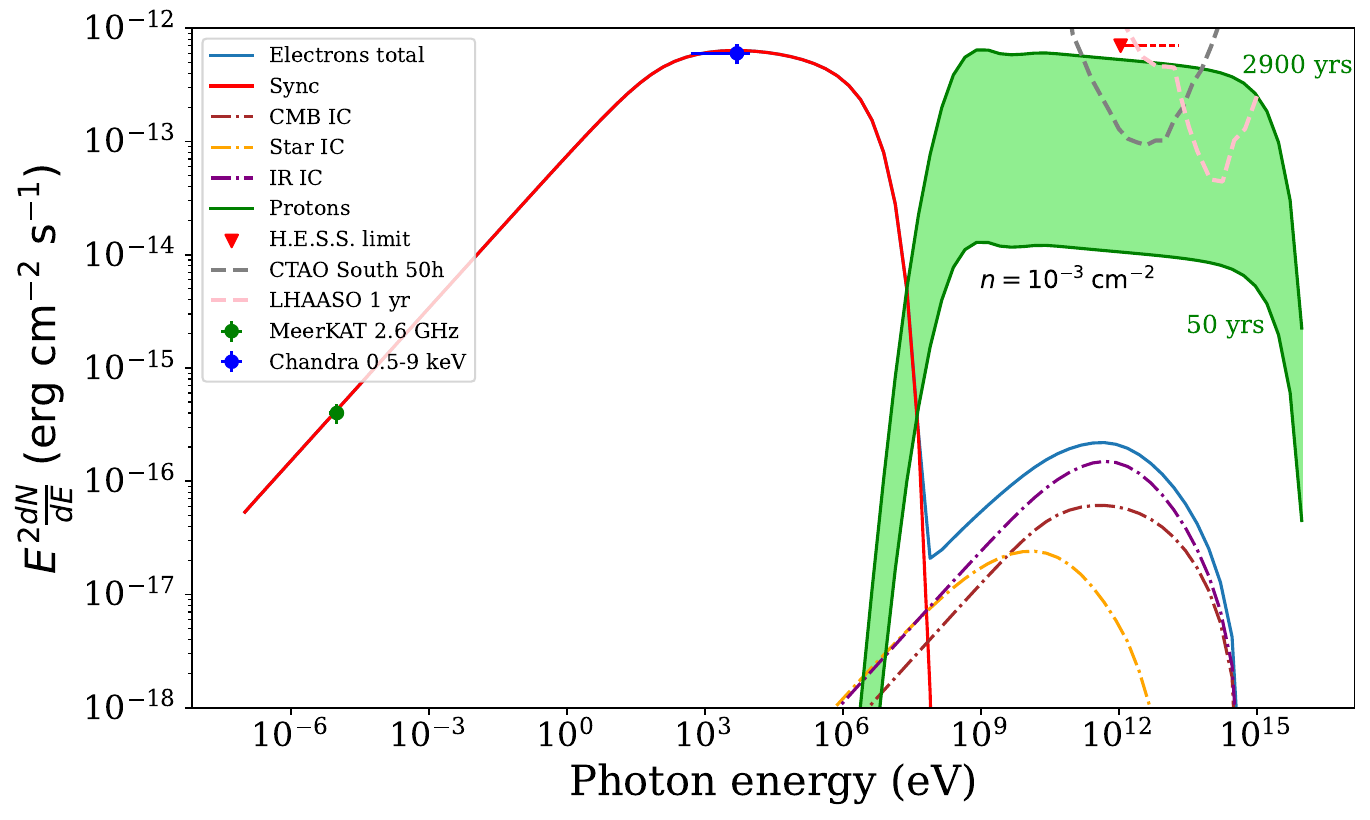}
 \caption{Model SED of the emission related to the shocks of Cir X-1. At lower photon energies the synchrotron contribution with a cooling break dominates. At higher photon energies, the dot-dashed lines show emission from inverse Compton of non-thermal electrons from various photon backgrounds, and the solid line shows their sum. However, in this model, in gamma-rays the flux is dominated by a hadronic component, which depends on the how long acceleration of high energy protons has been ongoing in the source. The shaded region is bounded by curves showing the 2 limiting cases, the lifetime of the shocks (50 years) and the system (2900 years). Also plotted are the MeerKAT and Chandra observed fluxes for the shocks, and an observed H.E.S.S. integral upper limit in the gamma-ray regime. The H.E.S.S. upper limit is for a point source at $0.07\degree$ resolution, so does not constrain the proton VHE emission which will come from a nearby area on the sky. We show the LHAASO 1 year sensitivity and predicted CTAO South 50 hour sensitivity curves as dashed lines.}
 \label{fig:vhe_model}
\end{figure*}

Given the recent detections of very high energy (VHE), $>100~\text{TeV}$, emission from numerous X-ray binaries (e.g. \citealt{lhaaso_2024}, \citealt{hess_2024}), including from Cyg X-1, GRS 1915, and SS 433 which all show large scale structure, we can attempt to predict the VHE emission for the shocks of Cir X-1. We use the \textsc{gamera} package \citep{gamera} to model a single zone of electrons and a separate zone of protons evolving in a time independent environment. We take into account losses due to inverse Compton, synchrotron, Bremsstrahlung, ionisation and adiabatic expansion for the electrons, and losses from synchrotron, inverse Compton and $pp$ interactions for the protons. We assume a time and energy constant escape time of $\frac{R}{u_s} = \frac{2~\text{pc}}{0.1c}$ for the electrons only, the same set of background radiation fields for both zones as was assumed for the cooling time calculations in Section~\ref{sec:particle_accel}, an ambient density of $10^{-2}~\text{cm}^{-3}$ within the SNR, and a $1~\text{pc}$ scale size with expansion at $0.05c$. 

For the protons we do not consider an escape time, as protons accelerated at the shock then diffuse out into the surrounding region. Using a value for the diffusion coefficient of $10^{30}~\text{cm}^2~\text{s}^{-1}$ for 1~PeV protons, which is consistent with estimates for escape time from the Galaxy \citep{wang_2025} we find that the protons will have travelled 30~pc away from Cir X-1 over the lifetime of the shocks. This opens up the possibility that the high energy protons will encounter a dense molecular cloud region where they can more efficiently interact and cool, leading to the production of VHE emission. This is especially relevant given the position of Cir X-1 at low galactic latitude ($l=0.04\degree$), where a higher density of molecular clouds are expected \citep{ballesteros_2020}. We find that in order to obtain any substantial VHE emission from the protons, a significantly higher density is required than the assumed ambient density of $10^{-2}~\text{cm}^{-3}$ in the SNR. Therefore, when calculating the VHE emission from protons we assume a density of $10^2~\text{cm}^{-3}$, typical for any molecular cloud the protons would interact with as they diffuse away from their acceleration site \citep{ballesteros_2020}.

We adopt an injection luminosity of electrons of $7\times10^{35}~\text{erg s}^{-1}$, and a magnetic field of $200~\mu\text{G}$, to be compatible with our combined flux measurements of both shocks and an age of $50~\text{years}$. We assume a low and high energy cut-off to the electron injection spectrum of $10~\text{MeV}$ and $0.7~\text{PeV}$. The low energy cut-off is within our observational constraints as discussed in Section~\ref{sec:energetics}, and changing it results in only minor changes to the resulting SED. The high energy cut-off results from our analysis in Section~\ref{sec:particle_accel}. The proton injection spectrum is set as $\eta\sim100$ times the luminosity of the electron injection spectrum. This value is commonly used, motivated by observations \citep{morlino_2012} and simulations \citep{gupta_2024}. We adopt a low and high energy cut-off to the proton injection spectrum of $2~\text{GeV}$ and $20~\text{PeV}$, where the low energy cut-off comes from a requirement that the protons be relativistic, and the high energy cut-off comes from the Hillas limit in Section~\ref{sec:particle_accel}. We then use the resulting particle spectra after 50 years of evolution to calculate, the synchrotron, inverse Compton, Bremsstrahlung and ionisation emission from the electrons, and the hadronic emission from the protons. The outflow powering the shocks may have been active for longer than the lifetime of the currently observed shocks, or if the shocks were produced by a more transient explosion, this may be a repeating event. These scenarios are discussed further in Section~\ref{sec:outflow_nature}, but in either case, while the emission from the non-thermal electron population is not expected to be affected due to synchrotron and adiabatic cooling, the emission from the high energy protons, which do not suffer such effects, may be altered. Assuming the same constant injection luminosity of high energy protons over the entire lifetime of Cir X-1 of $\sim2900$ years leads to a factor $\sim50$ increase in the hadronic VHE flux.

The overall resulting SED at a distance of $9.4~\text{kpc}$, broken into relevant components, is shown in Figure~\ref{fig:vhe_model}. The two curves which dominate at VHE energies represent the emission from a population of protons generated over 50 years (the shock lifetime) and 2900 years (the system lifetime). The VHE emission from the protons would lie in the shaded area between these two curves. We find Bremsstrahlung and ionisation emission to be negligible compared to synchrotron and inverse Compton. The radio and X-ray measurements are shown over-plotted, alongside an integral upper limit on the flux above $1~\text{TeV}$ from Cir X-1 \citep{hess_2018}. This upper limit is a point source upper limit (at the $\sim0.07\degree$ resolution of H.E.S.S.) on emission from Cir X-1 itself, and therefore does not constrain hadronic emission from a nearby over dense region as assumed in our model, but does applies to the leptonic VHE emission predicted from Cir X-1. The result is that there are currently no constraints on hadronic emission from Cir X-1 and further observations are needed. We stress that this SED is not a fit of a model to the available data, rather we have inferred the physical parameters as outlined in the preceding sections and used them to predict the SED. 

From this predicted SED we see first that the radio and X-ray fluxes are in good agreement with a synchrotron model. Secondly, we note that in the VHE regime, hadronic emission from a molecular cloud environment dominates over the inverse Compton emission from the non-thermal electron population due to the high magnetic field measured for the shocks. The majority of the parameter space for the hadronic VHE flux is below the levels detected in some other XRBs at similar distances. This is not unexpected given the youth of the system compared to other XRBs, meaning that perhaps a large population of high energy particles has not had sufficient time to build up. In the more optimistic cases where high energy protons have been accelerated for a significant fraction of the lifetime of Cir X-1 then the predicted flux may be detectable with next-generation gamma-ray observatories under construction such as CTA South (\citealt{cta_2013}, \citealt{bernlohr_2013}) or a similar instrument to LHAASO in the southern hemisphere such as SWGO (\citealt{lhaaso_book}, \citealt{swgo_paper}). This is demonstrated by the sensitivity curves in Figure~\ref{fig:vhe_model}. This highlights Cir X-1 as a promising candidate for the detection of VHE emission, in particular hadronic emission, in the future.

\subsection{Comparison to outflow interaction regions in other XRBs}\label{sec:comparison}

We can compare the large scale outflow interaction structure in Cir X-1 to those found in other XRBs. In almost all other XRBs the parsec scale structures observed seem to conform to the model of a radio lobe plus bow shock, powered by expanding hot gas, which is in turn heated by a steady collimated hard state jet \citep{kaiser_2004}. A similar model is used to explain the lobes of radio galaxies \citep{kaiser_1997}. This model is consistent with observations of Cyg X-1 (\citealt{gallo_2005}, \citealt{atri_2025}), GRS 1915+105 \citep{motta_2025}, GRS 1758-258 \citep{mariani_2025}, S26 (\citealt{pakull_2010}, \citealt{soria_2010}) and S10 (\citealt{urquhart_2019}, \citealt{mcleod_2019}). In all of these cases, the inferred kinematics indicate shock velocities of a few 100s of $\text{km s}^{-1}$ and lifetimes of $\gtrsim10~\text{kyr}$. Morphologically, the radio emission exhibits curved bow-shock structures produced by synchrotron or bremsstrahlung processes, and in some instances thermal X-ray emission is also detected; however, it is not spatially coincident with the radio structures (see \citealt{soria_2010} for a discussion of its origin).

The properties of the large scale shocks in Cir X-1 are in complete contrast to all of these. We measure a shock velocity two orders of magnitude larger than in these other sources, the lifetime of the shocks is also two orders of magnitude lower. No curved bow shock structure is seen, instead the shocks appear to have a constant opening angle and little curvature. Finally, X-ray synchrotron emission coincident with the radio emission is present, seen in no other sources. This reinforces that the shocks seen in Cir X-1 are not the analogue of those seen in other sources, and cannot be interpreted with the same model. Furthermore, it is not surprising that given the high velocity, short lifetime, and X-ray synchrotron emission, we calculate such large power requirements for the shocks.

The only other XRB system which does not fit into the bow shock plus lobe model is SS 433 \citep{margon_1984}, which also shares a remarkable number of other similarities with Cir X-1, being a peculiar XRB, residing inside a natal SNR where outflows have inflated bubbles (\citealt{downes_1986}, \citealt{goodall_2011b}, \textcolor{blue}{Gasealahwe, Savard et al. 2025}), launching precessing jets (\citealt{margon_1984}, \citealt{cowie_2025}), and having numerous features which could be identified as sites of outflow interaction. However, proper motion searches in both radio \citep{goodall_2011} and X-rays \citep{tsuji_2025} of these regions in SS 433 have revealed no significant proper motion down to limits of $\sim 0.03c$. This demonstrates a significant difference between the structures in Cir X-1 and in SS 433, and perhaps arises due to the different ages of the systems ($\sim2\times10^4~\text{years}$ for SS 433 \citealt{goodall_2011b} and $\sim3\times10^{3}~\text{years}$ for Cir X-1 \citealt{heinz_2013}).

Therefore, it is apparent from comparisons to outflow interaction structures in other XRBs that the shocks of Cir X-1 have several unique properties. Given the peculiarity of Cir X-1 as the youngest known XRB, and its extreme properties across the electromagnetic spectrum this is perhaps expected.

\subsection{Nature of the powering outflow}\label{sec:outflow_nature}

Given that in the shocks of Cir X-1 show so many differences when compared to other known XRB outflow interaction regions, and have extreme energetic properties, we briefly explore the possible nature and powering of the outflow responsible for these shocks. In Section~\ref{sec:geometry} we concluded that a wide conical outflow with a velocity of at least $\sim0.3c$ was the most likely explanation for shocks. Such an outflow has no immediate analogue with known outflows in XRBs. It is not collimated enough to be a jet, and too fast to be a typical wind. A potential explanation for the shocks is that they are powered by an ultra-fast outflow (UFO). These are fast ($\sim0.05c-0.3c$) outflows of highly ionized gas which are not highly collimated. They are observed in around half of all AGN \citep{tombesi_2011}, and some evidence has been found for their presence in both black hole XRBs (\citealt{del_santo_2023}, \citealt{miller_2025}) and neutron star XRBs \citep{eijnden_2019}, as well as in ultra-luminous X-ray sources \citep{pinto_2017}. The mechanism behind UFOs remains undetermined, and Cir X-1 may provide a site to study the interaction properties of such an outflow with the surrounding medium.

It is also possible that the powering outflow of the shocks is not continuous over their $\sim50~\text{year}$ lifetime, and instead the shocks were formed by a much more transient, explosive, event. Given the long term secular variation in X-ray and radio luminosity of Cir X-1 over several orders of magnitude since its discovery, and the young age of the system, there is no \textit{a priori} reason to assume a steady state. However, reducing the energy injection timescale requires higher instantaneous power further above the Eddington luminosity for a neutron star. Furthermore, if the outflow power is significantly variable over the shock lifetime we can imagine this may impact the speed of the shocks over time, leading to variable shock speed over time. In the case where the outflow is truly a transient event which occurred $\sim50$ years ago we can imagine exotic scenarios where the neutron star undergoes a large burst of hyper-Eddington accretion, e.g. accreting a large gas cloud or exoplanet, which briefly powers an outflow of material. 


Whether the outflow is continuous or transient, the analysis in Section~\ref{sec:energetics} reveals a vast amount of energy required to produce the shocks of Cir X-1. We can compare this to the apparent accretion and radio flare power of the source. While Cir X-1 does sometimes exceed the Eddington luminosity for a $1.4~\text{M}_\odot$ neutron star in X-rays it appears to have spent the majority of the past 50 years at sub-Eddington luminosities \citep{armstrong_2013}. The radio luminosity of Cir X-1 has varied by 3 orders of magnitude over 50 years, and therefore so to do estimates of the energy of these radio flares (\citealt{fender_bright_2019}, Cowie \& Fender in prep.). It is therefore not easy to explain an outflow requiring a significantly super-Eddington power given the accretion luminosity appears to be mostly sub-Eddington. This could be reconciled by high degrees of obscuration limiting the observed X-ray luminosity, making Cir X-1 an obscured ULX. This scenario is also invoked to explain the powerful jets of SS 433, where the observed X-ray luminosity of the XRB is also insufficient \citep{fabrika_2015}.


\section{Conclusions}



By combining archival data going back 24 years, with detailed new observations of the neutron star XRB Cir X-1, we have revealed dynamic shocks where a wide conical outflow, launched in the mid-20th century, interacts with the surrounding medium. Measuring the position over time of the shocks let us infer both their velocity and lifetime, allowing for the investigation of a wide variety of physics not typically possible in similar systems due to a lack of information. The shocks of Cir X-1 now represent one of the best studied examples of a parsec scale interaction region between an XRB powered outflow and the ambient medium. 

In Section~\ref{sec:cap_nature} we have robustly shown the shock nature of the structures and a synchrotron origin for the emission using their surface brightness profiles, high linear polarisation fraction of up to $\sim50\%$, and the spectral index of the radio and X-ray emission. In Section~\ref{sec:geometry} we demonstrate that the outflow powering the shocks must be inclined to the line of sight at $\gtrsim35\degree$, and is likely a wide conical outflow with a half opening angle $>17\degree$. This wide outflow must be mildly relativistic with $\beta \gtrsim 0.3c$, and as discussed in Section~\ref{sec:outflow_nature} has no obvious analogue with typical winds or jets in XRBs, instead potentially being better characterised as an ultra-fast outflow. We examined the energetics of this outflow in Section~\ref{sec:energetics} where our measured lifetime of the shocks of $\sim50$ years allowed for a determination of a robust lower limit on the time-averaged power required to produce the observed synchrotron radiation of $\sim10\%~L_\text{Edd}$. Furthermore, using the synchrotron cooling break combined with our measured lifetime of the shocks we obtained an independent estimate of the magnetic field in the shocks, higher than expected from the equipartition estimate, and one explanation for this is the presence of non-thermal protons. Our measured velocity for the shocks also allowed us to estimate the large total kinetic energy of the swept up material of $1\times10^{49}~\text{erg}$, requiring a time-averaged power of $7\times10^{39}~\text{erg~s}^{-1}\approx40 ~L_\text{Edd}$. 

Using our detailed knowledge of the physical conditions of the shocks we were then able to thoroughly investigate the particle acceleration potential of this environment in Section~\ref{sec:particle_accel}. The observation of synchrotron radiation up to X-rays in a $\sim200~\mu\text{G}$ magnetic field already requires electrons of energies of $\gtrsim30~\text{TeV}$. Knowledge of the shock velocity allowed us to calculate the expected acceleration timescale in the DSA framework and balancing this with the synchrotron cooling timescale we found electrons should be accelerated up to $\sim0.7~\text{PeV}$. The shock velocity also allowed us to calculate the Hillas limit, demonstrating the shocks are capable of accelerating protons up to $\sim20~\text{PeV}$. Using our energetic estimates for the shocks we are also able to estimate the energetic and injection efficiencies of accelerated electrons. Demonstrating that in future studies, XRBs and their outflows are a promising laboratory for studying particle acceleration in mildly relativistic shocks. 

Given the potential of the shocks to accelerate particles up to $10$s of PeV, in Section~\ref{sec:sed} we modelled the full SED of the shocks and examined the expected gamma-ray signature. We found that the gamma ray signature from the shocks is only detectable if there is a hadronic component arising from high energy protons interacting with a molecular cloud or other dense region after diffusing away from the acceleration site. Leptonic contributions to the gamma-ray signal are suppressed due to the high magnetic field in the shock. The hadronic signature may be detectable by next generation gamma-ray observatories allowing for unambiguous identification of gamma-rays from high energy protons.  

Overall, it is clear that the shocks of Cir X-1 have extreme properties that make them unique even among other XRB outflow interaction structures (Section~\ref{sec:comparison}). Detailed analysis of them has allowed for insights into outflow geometry and power, and the large impact outflows can have on the ambient medium. Future studies of this unique object promise more insights into XRB outflows and particle acceleration.

\section*{Acknowledgements}

FJC was supported by STFC grant ST/Y509474/1. RF acknowledges support from UK Research and Innovation, the European Research Council and the Hintze Charitable Foundation. JvdE was supported by funding from the European Union's Horizon Europe research and innovation programme under the Marie Skłodowska-Curie grant agreement No 101148693 (MeerSHOCKS). The MeerKAT telescope is operated by the South African Radio Astronomy Observatory, which is a facility of the National Research Foundation, an agency of the Department of Science and Innovation. This work has made use of the “MPIfR S-band receiver system” designed, constructed and maintained by funding of the MPI für Radioastronomy and the Max-Planck-Society. We acknowledge the use of the ilifu cloud computing facility – www.ilifu.ac.za, a partnership between the University of Cape Town, the University of the Western Cape, Stellenbosch University, Sol Plaatje University and the Cape Peninsula University of Technology. The ilifu facility is supported by contributions from the Inter-University Institute for Data Intensive Astronomy (IDIA – a partnership between the University of Cape Town, the University of Pretoria and the University of the Western Cape), the Computational Biology division at UCT and the Data Intensive Research Initiative of South Africa (DIRISA). This work made use of the CARTA (Cube Analysis and Rendering Tool for Astronomy) software (DOI 10.5281/zen- odo.3377984 – https://cartavis.github.io). The authors would like to thank Lilia Tremou, Andrew Hughes, Francesco Carotenuto and Payaswini Saikia for scheduling the MeerKAT observations. The Australia Telescope Compact Array is part of the Australia Telescope National Facility (https://ror.org/05qajvd42) which is funded by the Australian Government for operation as a National Facility managed by CSIRO. We acknowledge the Gomeroi people as the Traditional Owners of the Observatory site. This paper includes archived data obtained through the Australia Telescope Online Archive (http://atoa.atnf.csiro.au).

\section*{Data Availability}

The older data for this study is publicly available on the MeerKAT and ATCA archives. The data availability of more recent data is subject to the ThunderKAT LSP and X-KAT XLP data release conditions.



\bibliographystyle{mnras}
\bibliography{shocksbib} 




\clearpage
\appendix

\section{Polarisation intensity maps}\label{sec:pol_inten}

\begin{center}
 \includegraphics[width=\columnwidth]{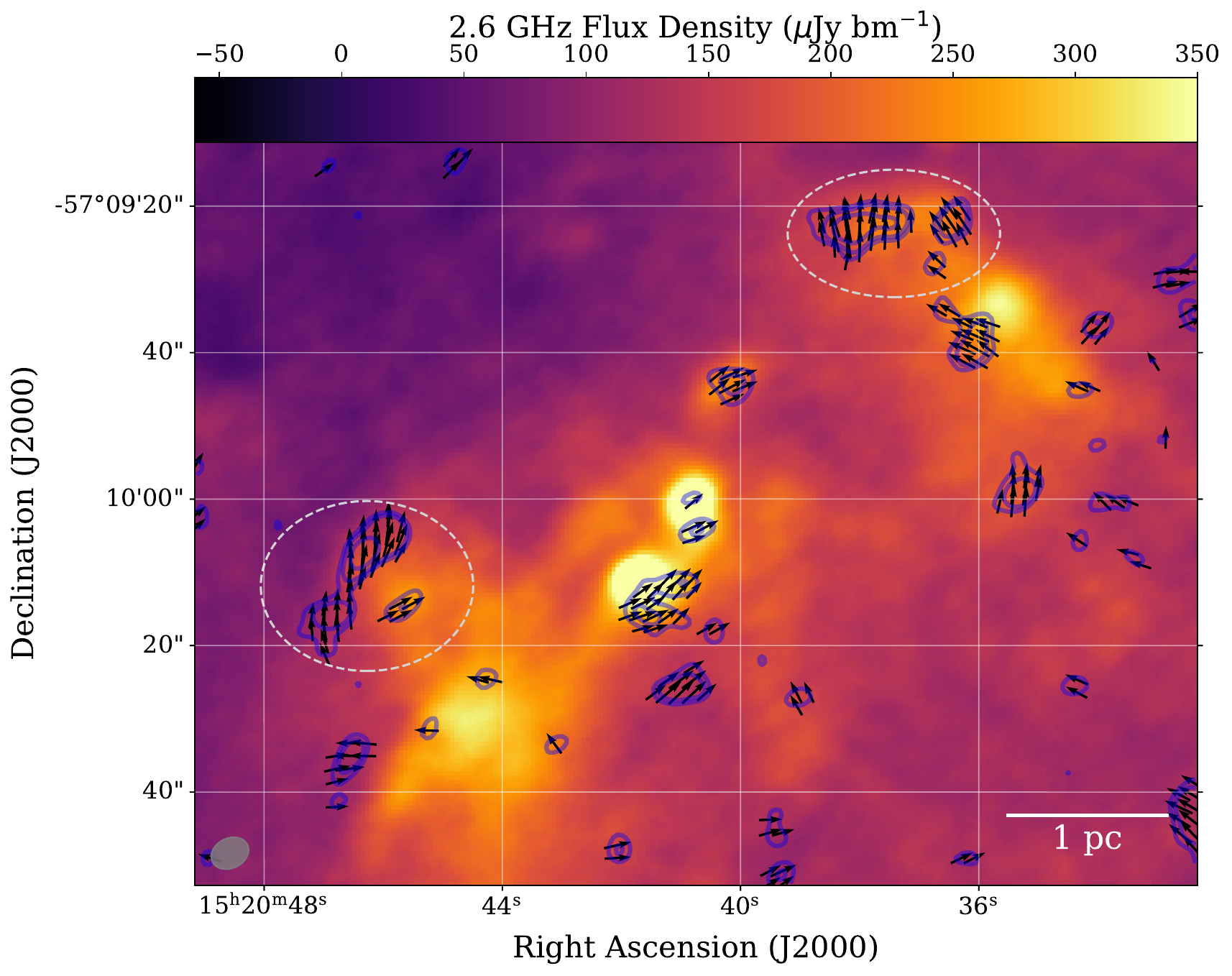}
 \captionof{figure}{Total intensity image of Cir X-1 and the surrounding shocks at 2.6 GHz with contours of $(4\sigma,5\sigma,7\sigma)$ for the maximum polarisation intensity over all Faraday depth over-plotted. $\sigma=3~\mu\text{Jy}$ and polarisation intensity values have been Ricean de-biased using $P=\sqrt{P_\text{obs}^2-2.3\sigma^2}$. All of the maximum polarisation intensities occur between Faraday depths of $\pm400~\text{rad m}^{-2}$ with mean Faraday depth of $\sim0~\text{rad m}^{-2}$ . Over-plotted are vector arrows showing the de-rotated polarisation angle or EVPA. The dashed ellipses mark regions where the highest levels of fractional linear polarisation were observed and measured as reported in Section~\ref{sec:results}. Fractional linear polarisation was measured over a solid angle the size of the beam (shown as an ellipse in the bottom left), so as to avoid any resolution effects. As well as the shocks, in linear polarisation we detect the southern component of the north-south jet close to the core, and tentatively detect a northern component. We also detect the northern and southern components of the S-shaped precessing jet.}
 \label{fig:shock_pol}
\end{center}

\section{Estimating physical parameters from synchrotron radiation}\label{sec:energetics_appen}

The minimum energy contained within a synchrotron emitting plasma of known volume, $Vf$, can be determined \citep{burbidge_1956} (where $V$ is the observed volume of emission and $f$ is the filling factor of the plasma in this volume). We follow the treatment outlined in \cite{pacholczyk} in order to do this. Synchrotron emission is due to a population of non-thermal electrons, assumed to follow a power-law distribution in energy, with a low and high energy cut-off ($E_\text{min}$ and $E_\text{max}$) as expected from acceleration mechanisms (e.g. \citealt{bell_1978}):

\begin{equation} \label{eq:electron_distribution}
    N(E) = N_0 E^{-p} , 
\end{equation}

\noindent where $N$ the density density of electrons (spatial density and energetic density). $E$ is the electron energy, and $N_0$ is a normalisation with units depending on $p$. The energy in a synchrotron emitting plasma is contained in these non-thermal particles:

\begin{equation} \label{eq:electron_energy}
    E_e = fV\int^{E_\text{max}}_{E_\text{min}} EN(E)dE = fV \frac{N_0}{2-p} (E_\text{max}^{2-p} - E_\text{min}^{2-p}) ,
\end{equation}

\noindent and within the magnetic field, $B$:

\begin{equation} \label{eq:mag_field_energy}
    E_B = fV \frac{B^2}{8\pi} .
\end{equation}

\noindent We can then relate the observed optically thin luminosity, $L$, of the synchrotron emission (e.g. the intrinsic luminosity neglecting (self)absorption effects) to the energy in electrons by using the radiated synchrotron power of an electron in a magnetic field:

\begin{equation}
    -\frac{dE}{dt} = c_2 B^2 E^2 , 
\end{equation}

\noindent where $c_2$ is a constant given by:

\begin{equation}
    c_2 = \frac{2e^4}{3 m_e^4 c^7} .
\end{equation}

\noindent Calculating the synchrotron luminosity we obtain:

\begin{align} \label{eq:theretical_lum}
    L & = -fV\int^{E_\text{max}}_{E_\text{min}} \frac{dE}{dt} N(E) dE \nonumber \\ & = fV N_0 c_2 B^2 \int^{E_\text{min}}_{E_\text{max}} E^{2-p} dE \nonumber \\ & = fV\frac{N_0 c_2 B^2}{3-p}(E_\text{max}^{3-p}-E_\text{min}^{3-p}) .
\end{align}

\noindent Combining this with the equation for the characteristic emission frequency of an electron of a given energy:

\begin{equation} \label{eq:crit_freq}
    \nu_\text{cr} = c_1BE^2 , 
\end{equation}

\noindent where $c_1$ is a constant given by:

\begin{equation}
    c_1 = \frac{3e}{4\pi m_e^3c^5} .
\end{equation}

\noindent We can eliminate $N_0$ using equations \eqref{eq:electron_energy} and \eqref{eq:theretical_lum} to obtain:

\begin{equation}\label{ee_eqn}
    E_e = c_{12}(p,\nu_\text{min},\nu_\text{max})LB^{-\frac{3}{2}} ,
\end{equation}

\noindent where $c_{12}$ is a constant depending on given by:

\begin{equation}
    c_{12} = c_2^{-1} c_1^{1/2} \frac{(p-3)}{(p-2)} \frac{\nu_\text{max}^{(2-p)/2}-\nu_\text{min}^{(2-p)/2}}{\nu_\text{max}^{(3-p)/2}-\nu_\text{min}^{(3-p)/2}} .
\end{equation}

\noindent The total energy in the synchrotron plasma is then:

\begin{equation}\label{eq:total_energy}
    E_\text{tot} = E_B + (1+\eta) E_e \propto B^2 + B^{-\frac{3}{2}} , 
\end{equation}

\noindent where $\eta$ is the ratio between energy stored in non-thermal protons and non-thermal electrons. For the purposes of a minimum energy estimate, it is assumed the plasma only contains non-thermal electrons, so $\eta=0$. From equation~\eqref{eq:total_energy} we can see that given a luminosity and emitting volume but unknown magnetic field, there exists a minimum in the energy close to equipartition, when:

\begin{equation}\label{eq:min_energy_condition}
    E_B = \frac{3}{4} (1+\eta)E_e ,
\end{equation}

\noindent or equivalently:

\begin{equation} \label{eq:equip_energy_in_terms_of_eb}
    E_\text{tot} = \frac{7+3\eta}{3(1+\eta)} E_B .
\end{equation}

\noindent We can use this, combined with equations~\eqref{eq:mag_field_energy} and \eqref{ee_eqn} above in order to find the minimum energy associated with a synchrotron emitting plasma, and the associated equipartition magnetic field. We calculate the optically thin luminosity of the synchrotron emission using our observation of the flue density at a given frequency, $F_{\nu_\text{obs}}$, the spectral index of the emission, $\alpha=\frac{1-p}{2}$, the distance to the source $D$, and the frequency range spanned by the synchrotron emission, between $\nu_\text{min}$ and $\nu_\text{max}$:

\begin{align}
    L & = 4\pi D^2 \int_{\nu_\text{min}}^{\nu_\text{max}} F_\nu d\nu \nonumber \\ & = 4\pi D^2 F_{\nu_\text{obs}} \nu_\text{obs}^{-\alpha} \left( \frac{\nu_2^{\alpha+1} - \nu_1^{\alpha+1}}{\alpha+1} \right) . 
\end{align}

\noindent Finally, the minimum energy and associated magnetic field are given by:

\begin{align}
    B_\text{eq}  & = \left(\frac{6 \pi (1+\eta)}{fV}c_{12} L\right)^\frac{2}{7} \nonumber \\ & = \left(\frac{6 \pi (1+\eta)}{fV} \frac{2 c_2^{-1} c_1^{\frac{1}{2}}}{1+2\alpha}   4\pi D^2 F_{\nu_\text{obs}} \nu_\text{obs}^{-\alpha}\left(\nu_\text{max}^\frac{1+2\alpha}{2}-\nu_\text{min}^\frac{1+2\alpha}{2}\right)\right)^\frac{2}{7} , 
\end{align}

\begin{align}
    E_\text{min} & = \frac{7}{4}(1+\eta) (c_{12} L)^\frac{4}{7} \left(\frac{6 \pi (1+\eta)}{fV} \right)^{-\frac{3}{7}} \nonumber \\ & = \frac{7}{4}(1+\eta)^{\frac{4}{7}} \left(\frac{6 \pi}{fV} \right)^{-\frac{3}{7}} \nonumber  \\ & \cdot \left( \frac{2 c_2^{-1} c_1^{\frac{1}{2}}}{1+2\alpha}   4\pi D^2 F_{\nu_\text{obs}} \nu_\text{obs}^{-\alpha}\left(\nu_\text{max}^\frac{1+2\alpha}{2}-\nu_\text{min}^\frac{1+2\alpha}{2}\right) \right)^\frac{4}{7} ,
\end{align}

\noindent where we have made explicit the dependence on $\nu_\text{min}$ and $\nu_\text{max}$ in $L$ and $c_{12}$. The total number of electrons can also be calculated by integrating \eqref{eq:electron_distribution}:

\begin{equation}
    N_e = \frac{4}{7}{E_\text{tot}} \frac{1+2\alpha}{2\alpha} c_1^\frac{3}{2}B^\frac{3}{2} \frac{\left(\nu_\text{max}^\alpha - \nu_\text{min}^\alpha\right)}{\left(\nu_\text{max}^\frac{1+2\alpha}{2} - \nu_\text{min}^\frac{1+2\alpha}{2}\right)} .
\end{equation}

\begin{figure}
 \includegraphics[width=\columnwidth]{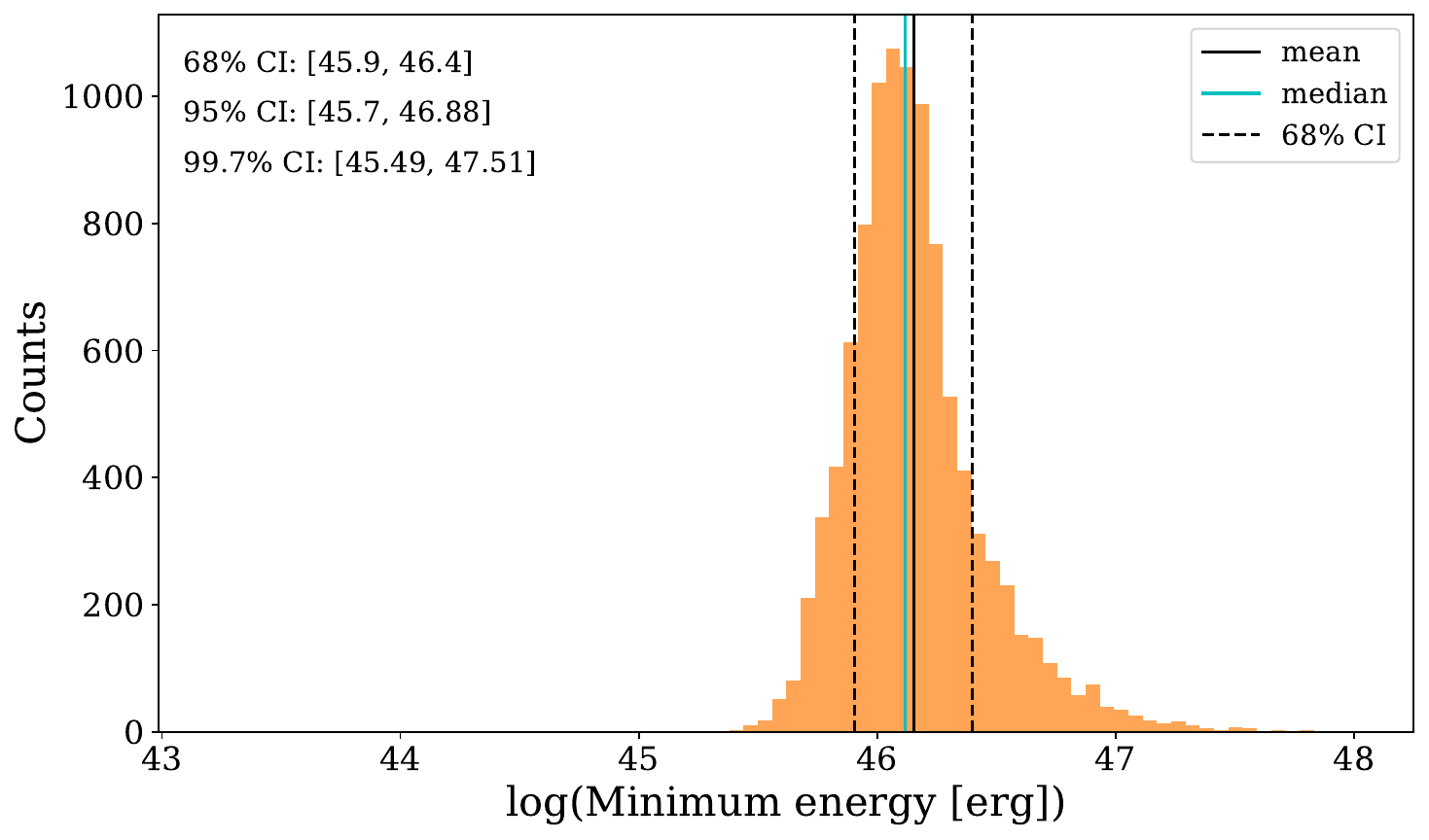}
 \caption{Posterior distribution for $E_\text{min}$ made of $10^4$ samples. The intervals in which 68\%, 95\%, and 99.7\% of the samples lie are shown in the top left of the figure.}
 \label{fig:energy_post}
\end{figure}

\begin{figure}
 \includegraphics[width=\columnwidth]{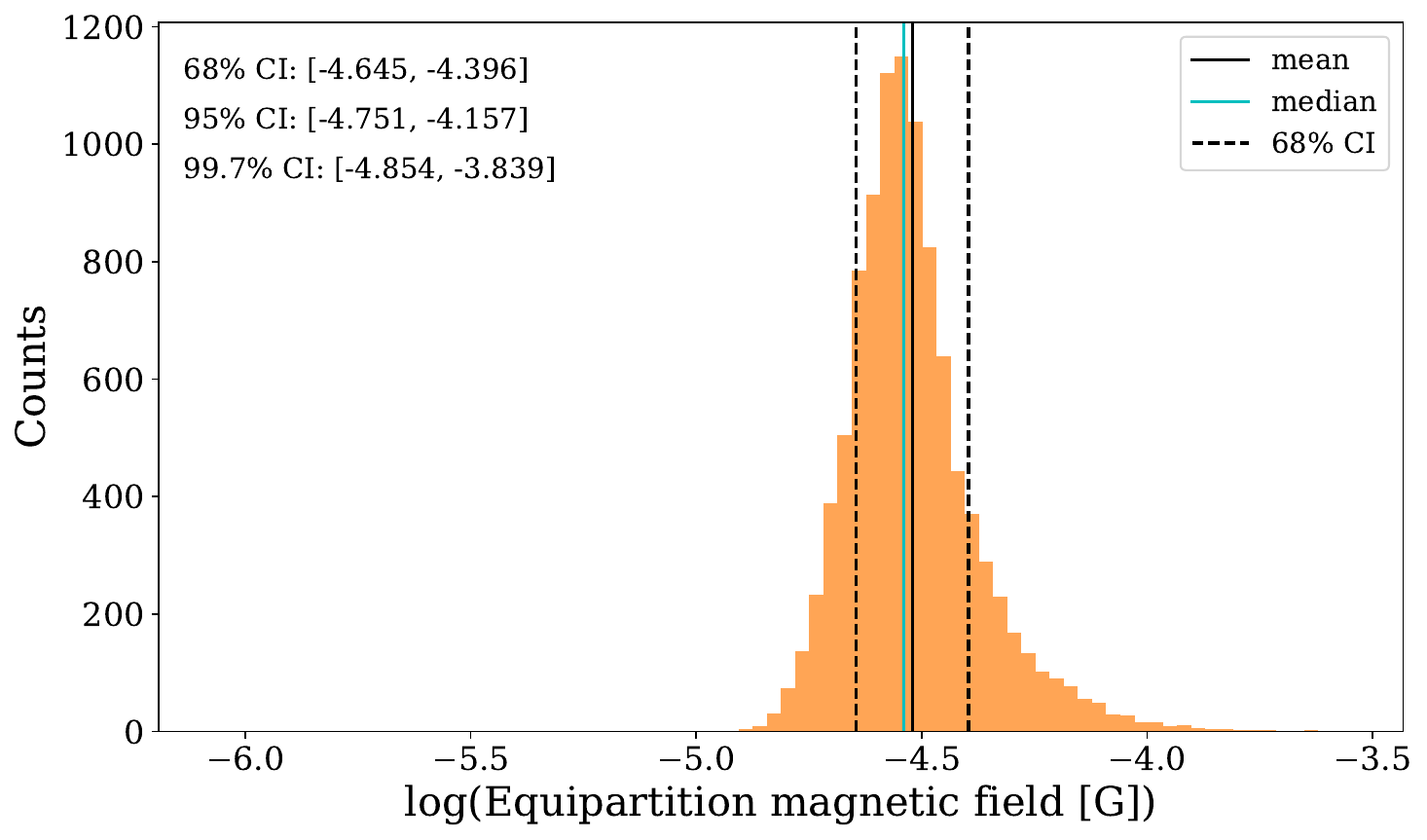}
 \caption{Posterior distribution for $B_\text{eq}$ made of $10^4$ samples. The intervals in which 68\%, 95\%, and 99.7\% of the samples lie are shown in the top left of the figure.}
 \label{fig:mag_post}
\end{figure}

\begin{figure}
 \includegraphics[width=\columnwidth]{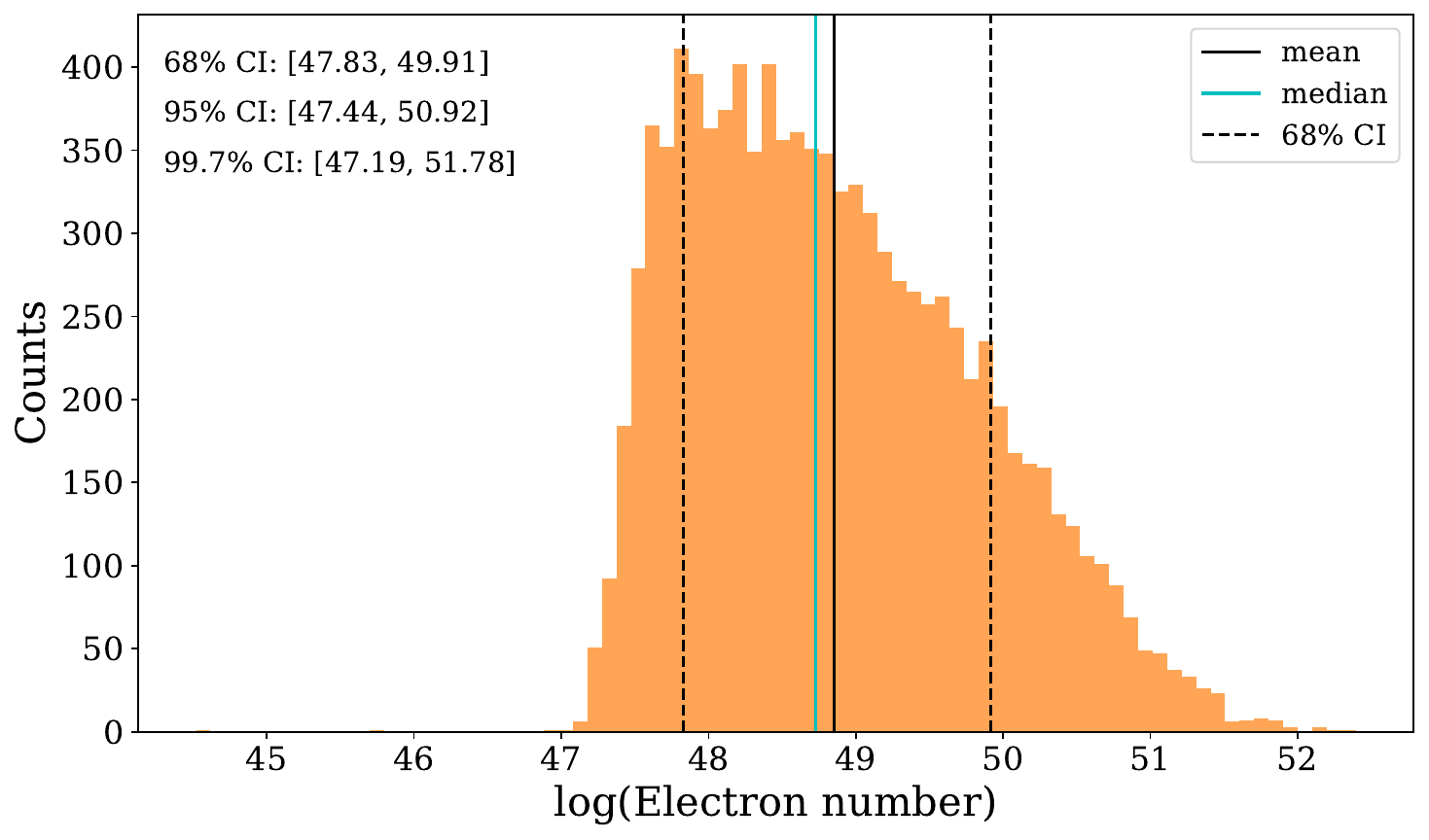}
\caption{Posterior distribution for $N_e$ made of $10^4$ samples. The intervals in which 68\%, 95\%, and 99.7\% of the samples lie are shown in the top left of the figure.}
 \label{fig:Ne_post}
\end{figure}

\noindent Using our observed parameters, outline in Section~\ref{sec:energetics}, we run Monte Carlo sampling on $10^4$ samples from our observed parameter priors, to generate posterior distributions for $E_\text{min}$, $B_\text{eq}$, and $N_e$, shown in Figures~\ref{fig:energy_post} to \ref{fig:Ne_post}.

\bsp	
\label{lastpage}
\end{document}